\documentclass[10pt]{iopart}
\usepackage{latexsym}
\usepackage{amssymb}
\usepackage{cite}

\begin{document}

\newcommand{\be}{\begin{equation}}
\newcommand{\ee}{\end{equation}}

\newcommand{\beqn}{\begin{eqnarray}}
\newcommand{\eeqn}{\end{eqnarray}}

\setlength{\arraycolsep}{2pt}

\newcommand{\bea}{\begin{eqnarray}}
\newcommand{\eea}{\end{eqnarray}}

\newcommand{\pa}{\partial}
\newcommand{\pp}{{\it pp\,}-}
\newcommand{\ba}{\begin{array}}
\newcommand{\ea}{\end{array}}

\def \Mm {\stackrel{m}{M}}
\def \Mmm {\stackrel{2m}{M}}

\def \eq5d {\stackrel{{\mbox{\tiny (n=5)}}}{=}}
\def \kd #1 {\delta_{#1}}
\def \beah {\begin{eqnarray*}}
\def \eeah {\end{eqnarray*}}

\def \bl {\mbox{\boldmath{$\ell$}}}
\def \bn {\mbox{\boldmath{$n$}}}
\def \bm #1 {\mbox{\boldmath{$m^{(#1)}$}}}
\def \bmd #1 {\mbox{\boldmath{$m_{(#1)}$}}}
\def \bk {\mbox{\boldmath{$k$}}}
\def \pul {{{\textstyle{\frac{1}{2}}}}}
\def \WDS #1 {\mbox{$\Phi_{#1}^{S}$}}
\def \WDA #1 {\mbox{$\Phi_{#1}^{A}$}}
\def \WD #1 {\mbox{$\Phi_{#1}$}}
\def \ble {\mbox{\boldmath{$e$}}}

\def \mS {\mbox{\boldmath{$S$}}}
\def \mA {\mbox{\boldmath{$A$}}}
 \def \mL  {\mbox{\boldmath{$L$}}}
\def \mPhi {\mathbf \Phi}

\def \mPhiS {\mathbf \Phi^S}
\def \mPhiA {\mathbf \Phi^A}

\def \d {{\rm d}}

\def \Mi {\stackrel{i}{M}}
\def \Mj {\stackrel{j}{M}}
\def \Mk {\stackrel{k}{M}}
\def \Mr {\stackrel{r}{M}}
\def \Ms {\stackrel{s}{M}}
\def \Ma {\stackrel{a}{M}}

\newcommand{\fvec}[4]{{#1}_{\{a}{#2}_b{#3}_c{#4}_{d\}}}
\newcommand{\nlnm}[1]{\fvec{n}{\ell}{n}{m^{(#1) \! \! }}}
\newcommand{\nmnm}[2]{\fvec{n}{m^{(#1) \! \! }}{n}{m^{(#2) \! \! }}}
\newcommand{\nllm}[1]{\fvec{n}{\ell}{\,\ell}{m^{(#1) \! \! }}}
\newcommand{\lnlm}[1]{\fvec{\ell}{n}{\ell}{m^{(#1)}}}
\newcommand{\lmlm}[2]{\fvec{\ell}{m^{(#1) \! \! }}{\,\ell}{m^{(#2) \! \! }}}
\newcommand{\nmlm}[2]{\fvec{n}{m^{(#1) \! \! }}{\ell}{m^{(#2) \! \! }}}
\newcommand{\nlnl}{\fvec{n}{\ell}{n}{\ell}}
\newcommand{\nlmm}[2]{\fvec{n}{\ell}{m^{(#1) \! \! }}{m^{(#2) \! \! }}}
\newcommand{\nmmm}[3]{\fvec{n}{m^{(#1) \! \! }}{m^{(#2) \! \! }}{m^{(#3) \! \! }}}
\newcommand{\lmmm}[3]{\fvec{\ell}{m^{(#1) \! \! }}{m^{(#2) \! \! }}{m^{(#3) \! \! }}}
\newcommand{\mmlm}[3]{\fvec{m^{(#1) \! \! )}}{m^{(#2) \! \! }}{\,\ell}{m^{(#3) \! \! }}}
\newcommand{\mmmm}[4]{\fvec{m^{(#1) \! \! }}{m^{(#2) \! \! }}{m^{(#3) \! \! }}{m^{(#4) \! \! }}}

\newcommand{\M}[3] {{\stackrel{#1}{M}}_{{#2}{#3}}}

\newtheorem{proposition}{Proposition}
\newtheorem{theorem}[proposition]{Theorem}
\newtheorem{lemma}[proposition]{Lemma}
\newtheorem{definition}[proposition]{Definition}
\newtheorem{remark}{Remark}

\title{Type D {Einstein} spacetimes in higher dimensions}

\author{V. Pravda$^1$, A. Pravdov\' a$^1$ and M. Ortaggio$^{2}$
\footnote{Now at: Departament de F\'{\i}sica Fonamental,
Universitat de Barcelona, Diagonal 647, E-08028 Barcelona, Spain}
}

\address{$^1$ Mathematical Institute,
Academy of Sciences, \v Zitn\' a 25, 115 67 Prague 1, Czech Republic}
\address{$^2$ Dipartimento di Fisica, Universit\`a degli Studi di Trento,
    and INFN, \\ Gruppo Collegato di Trento, Via Sommarive 14, 38050 Povo (Trento), Italy  }
\eads{\mailto{pravda@math.cas.cz}, \mailto{pravdova@math.cas.cz}, \mailto{ortaggio`AT'ffn.ub.es}}

\begin{abstract}
We show that all {\em static}  spacetimes in higher dimensions {$n>4$}  are necessarily of Weyl types G, ${\rm I}_i$, D {or O}. This applies also to {\em stationary} spacetimes provided  additional conditions  are fulfilled, as for most known black hole/ring solutions. (The conclusions change when the Killing generator becomes null, such as at Killing horizons, on which we briefly comment.) Next we demonstrate that the same Weyl types characterize {\em warped product spacetimes} with a one-dimensional
Lorentzian (timelike) factor, whereas warped spacetimes with a two-dimensional
Lorentzian factor are restricted to the types D or O.

{By exploring algebraic consequences of the Bianchi identities,} we then analyze the simplest {non-trivial} case from the above classes - type D vacuum
spacetimes, possibly with a cosmological constant, dropping, however, the assumptions that the spacetime is static, stationary or warped.
  It is shown that for ``generic'' type D vacuum  spacetimes (as defined in the text) the corresponding principal null directions
  are {\em geodetic} in arbitrary dimension (this in fact applies also to type II spacetimes).
For $n\geq 5$, however, there may exist particular cases of type D vacuum spacetimes which admit non-geodetic multiple principal null directions
 and we explicitly present such examples in any $n\ge 7$.

Further studies are restricted to five dimensions, where the {type D} Weyl tensor is fully described by a $3 \times 3$ real
matrix $\WD{ij} $. In the case with ``twistfree'' ($A_{ij}=0$)  principal null
geodesics we show that in a ``generic'' case $\WD{ij} $ is symmetric and eigenvectors of $\WD{ij} $ coincide with eigenvectors
of the expansion matrix $S_{ij}$ providing us thus in general with three preferred spacelike directions of the spacetime.
Similar results are also obtained when relaxing the twistfree condition and assuming instead that  $\WD{ij} $ is symmetric. {The five dimensional} Myers-Perry black hole and Kerr-NUT-AdS metrics in arbitrary dimension are also briefly studied as  specific {illustrative} examples of  type D vacuum spacetimes.
\end{abstract}


\section  {Introduction}

Algebraically special spacetimes play {an} essential role in the field of exact solutions of Einstein's equations
and many known exact solutions in four dimensions are {indeed} algebraically special \cite{Stephanibook}.
Recently a generalization of the Petrov classification to higher dimensions was developed
in \cite{Coleyetal04,Milsonetal05} and it
turned out that many higher-dimensional solutions of Einstein's equations are
algebraically special as well (see e.g. \cite{ColPel06}),
 in fact {so far} there is only one known solution identified \cite{OrtPra06} as algebraically general - the {static} charged
black ring \cite{IdaUch03}.

There is, however, one important difference between four dimensional and $n>4$ dimensional cases - the Goldberg-Sachs
theorem does not hold in higher dimensions. Recall that for $n=4$ the Goldberg-Sachs theorem implies that
principal null directions of an algebraically special vacuum spacetime are necessarily geodetic and shearfree.
It was stressed already in \cite{FroSto03,Pravdaetal04} that the Goldberg-Sachs theorem cannot be straightforwardly extended to higher dimensions.  Namely in  \cite{FroSto03} it was pointed out that principal null directions
(or Weyl aligned null directions - WANDs \cite{Coleyetal04})
of the $n=5$ Myers-Perry black holes \cite{MyePer86} are shearing though the spacetime is of type D.
In \cite{Pravdaetal04} it was shown that in fact all  vacuum, $n>4$, type N and III expanding spacetimes are shearing.
In \cite{OrtPraPra07}  it was also shown that for $n>4$, $n$ odd, all geodetic WANDs with non-vanishing twist are again shearing.

In this paper we study various properties of algebraically special
vacuum spacetimes, such as geodeticity of multiple WANDs (not
guaranteed in higher dimensions - another ``violation'' of the
Goldberg-Sachs theorem) and relationships between optical matrices
$S_{ij}$ and $A_{ij}$ and the Weyl tensor. Before approaching
these problems, we study in the first part of the paper (sections \ref{GID} and \ref{warped})
constraints on Weyl types of the spacetime following from
various  assumptions on the geometry.

In section   \ref{GID} we show that in arbitrary dimension (i.e.,
hereafter, $n\ge 4$) the only Weyl types compatible with static
spacetimes (and  expanding stationary spacetimes with appropriate
reflection symmetry) are types G, ${\rm I}_i$, D and O.

In section   \ref{warped} we study direct or warped product spacetimes.
It turns out that warped spacetimes with a one-dimensional Lorentzian
factor are again of types G, ${\rm I}_i$, D and O and
that warped spacetimes with a two-dimensional Lorentzian factor are
necessarily of type D or O. This also implies that spherically symmetric
spacetimes are of type D or O.

It follows  that type D
spacetimes play an important role as the simplest non-trivial case
compatible with the above mentioned assumptions.
Therefore, in the second part of the
paper (sections \ref{DHD} and \ref{D5D}) we focus on studying properties of type D {Einstein} spacetimes {(i.e., vacuum with an arbitrary cosmological constant)},
dropping, however, the assumptions that the spacetime is static, stationary or warped.

In section   \ref{DHD} we study type D spacetimes in arbitrary
dimension and analyze geodeticity of
WANDs. It turns out that in a ``generic''
case in vacuum the multiple WANDs are geodetic. Let us also point
out that  negative boost weight  Weyl components do not enter
relevant equations and thus the same results also hold
for multiple WANDs in type II Einstein spacetimes. Surprisingly, it also
turns out that explicit examples of special vacuum type D
spacetimes not belonging to our ``generic'' class and admitting
non-geodetic multiple WANDs  can easily be constructed. Such examples
for arbitrary dimension $n \geq 7$ are given in section   \ref{nongeod}.
This shows that
there exist even more striking ``violations'' of the
Goldberg-Sachs theorem in higher dimensions than {the} examples
{with non-zero shear}  discussed above. In section   \ref{DHD} we also
study various properties of shearfree type D vacuum spacetimes.

Perhaps not surprisingly,  the situation in five dimensions is
considerably simpler than for $n>5$. In fact it turns out that for
$n=5$ the  Weyl tensor of type D is fully determined by a $3
\times 3$ real matrix $\WD{ij} $. {At the same time, five dimensional gravity is already an interesting arena where qualitatively new phenomena appear.} We thus devote section   \ref{D5D} to
five dimensional vacuum type D spacetimes.  We  study
relationships between the Weyl tensor represented by  $\WD{ij} $
and optical matrices $S_{ij}$ and $A_{ij}$. One of the results is
that for  ``generic'' spacetimes with non-twisting WANDs
($A_{ij}=0$) the antisymmetric part of
 $\WD{ij} $, $\WDA{ij} $, vanishes and the symmetric part  $\WDS{ij} $ is aligned with $S_{ij}$ (in the sense that
 the matrices $\WDS{ij} $ and $S_{ij}$ can be diagonalized together). Similarly,
in the ``generic'' case {the condition}
 $\WDA{ij} =0 $ implies  vanishing of $A_{ij}$. Again, there exist particular cases for which the ``generic'' proof does not hold, see section   \ref{D5D}
 for details.
 In this section   a simple explicit example of a five-dimensional vacuum type D spacetime, the Myers-Perry metric, is also presented and  $S_{ij}$, $A_{ij}$, $\WDS{ij} $, and $\WDA{ij} $ are explicitly given.

Finally in  section   \ref{discussion}  we concisely summarize main results and in the Appendix we briefly study { geometric optics of} type D Kerr-NUT-AdS metrics in arbitrary dimension.

\label{sec_introduction}

\section  {Preliminaries}
\label{prelim}

Let us first briefly summarize our notation, further details can be found in \cite{Pravdaetal04}.
In an $n$-dimensional spacetime let us introduce a frame of $n$ real vectors $\bm{a} $ ($a, b\dots=0,\dots,n-1$):
two null vectors $\bm{0} =\bmd{1} =\bn$, $\bm{1} =\bmd{0} =\bl $ and $n-2$ orthonormal spacelike vectors
$\bm{i} =\bmd{i} $  ($i, j  \dots=2,\ldots,n-1$) satisfying
\be
\fl
\ell^a \ell_a= n^a n_a =\ell^a m^{(i)}_{a}=n^a m^{(i)}_a= 0, \qquad  \ell^a n_a = 1, \qquad
        m^{(i)a}m^{(j)}_a=\delta_{ij}.  \label{frame}
\ee
The metric now reads
\be
g_{a b} = 2\ell_{(a}n_{b)} + \delta_{ij} m^{(i)}_a m^{(j)}_b.
\ee
We will use the following decomposition of  the covariant derivative
of the vector $\bl $ and the covariant derivative in the direction of $\bl $
\be
\ell_{a;b}=L_{cd} m^{(c)}_a m^{(d)}_{b} \ , \qquad
\label{derlnm}
D \equiv \ell^a \nabla_a .
\ee

Note that $\bl$ is geodetic iff $L_{i0}=0$ and for an affine parameterization also $L_{10}=0$.
We will often use the symmetric and antisymmetric parts of $L_{ij}$,  $S_{ij}\equiv L_{(ij)}$
(its trace $S\equiv S_{ii}$), $A_{ij}\equiv L_{[ij]}$.
In case of geodetic $\bl$,  the trace of $S_{ij}$  represents expansion $\theta \equiv \frac{1}{n-2} S$,
the tracefree part of $S_{ij}$ is shear $\sigma_{ij} \equiv S_{ij} -\theta \delta_{ij}$ and the antisymmetric matrix
$A_{ij}$ is twist.\footnote{For the sake of brevity, throughout the paper we shall refer to the corresponding quantities for non-geodetic congruences as ``expansion'',
``shear'', and ``twist'' (in inverted commas), bearing in mind that in that case expressions (\ref{scalars}) do {\em not} hold.} Optical scalars can be expressed in terms of $\bl$ {(when $L_{i0}=0=L_{10}$)}
\be
\fl
  \sigma^2 \equiv \sigma_{ij} \sigma_{ji} =  \ell_{(a;b)}\ell^{(a;b)}-\textstyle{\frac{1}{n-2}}\left(\ell^a_{\;;a}\right)^2 ,
\ \ \  \theta = \textstyle{\frac{1}{n-2}}\ell^a_{\;;a} , \ \ \
 \omega^2\equiv A_{ij}A_{ij}=\ell_{[a;b]}\ell^{a;b} .
 \label{scalars}
\ee

The decomposition of the Weyl tensor in the frame (\ref{frame}) in full generality
is given by \cite{Pravdaetal04}
\beah
\fl
  C_{abcd} =
  %\overbrace{
    4 C_{0i0j}\, \nmnm{i}{j}
%}^2
  %\\
  \nonumber
  +%\overbrace{
    8C_{010i}\, \nlnm{i} +
    4C_{0ijk}\, \nmmm{i}{j}{k}
%}^1
 %+
  \nonumber \\
  \fl
  %\left\{
    %\begin{array}{l}
      \ \ \ \ \  \ \ \ \ \  {}+ 4 C_{0101}\, \nlnl +  4 C_{01ij}\, \nlmm{i}{j}+ %\\
      8 C_{0i1j}\, \nmlm{i}{j}
\\
\fl
\ \ \ \ \  \ \ \ \ \   {}+  C_{ijkl}\, \mmmm{i}{j}{k}{l}
    %\end{array}
    %\right\}^0 +
    %\overbrace{
   + 8 C_{101i}\, \ell_{\{a} n_b \ell_c m^{(i)}{\! \! }_{d\}}
 \nonumber \\
 \fl
  \ \ \ \ \  \ \ \ \ \ {}+
    4 C_{1ijk}\, \ell_{\{a} m^{(i)}{\! \! }_b m^{(j)}{\! \! }_c m^{(k)}{\! \! }_{d\}}
%}^{-1}
    + %\overbrace{
      4 C_{1i1j}\, \ell_{\{a} m^{(i)}{\! \! }_{b}  \ell_{c}  m^{(j)}{\! \! }_{b\}},
%}^{-2} ,
\eeah
where
the operation \{ \} is defined as
$w_{\{a} x_b y_c z_{d\}} \equiv \frac{1}{2}(w_{[a} x_{b]} y_{[c} z_{d]}+ w_{[c} x_{d]} y_{[a} z_{b]})$.

In the second part of this paper we will focus
on type D spacetimes, possessing (in an adapted frame) only boost order zero components (see \cite{Pravdaetal04})
$C_{0101}$, $C_{01ij}$, $C_{0i1j}$, $C_{ijkl}$.
For simplicity let us {define the $(n-2)\times (n-2)$ real
matrix}
\be
 \WD{ij} \equiv C_{0i1j} ,
\ee
 with $\WDS{ij} $, $\WDA{ij} $, and $\WD{ } \equiv \WD{ii} $ being the symmetric and antisymmetric parts
of $\WD{ij} $ and its trace, respectively. {Let us observe that for static spacetimes and for a large class of warped geometries one has $\WDA{ij} =0$ (see section   \ref{warped}).}
Note also that the above mentioned boost order zero components of the Weyl tensor
are not completely independent.
In fact from {the symmetries and the tracelessness of the Weyl tensor} (cf. eqs.~(7) and (9) in  \cite{Pravdaetal04}) it follows
that
\be
\fl
C_{01ij}=2 C_{0[i|1|j]}=2\WDA{ij} ,\ \quad C_{0(i|1|j)}=\WDS{ij} = - \textstyle{\frac{1}{2}} C_{ikjk}, \ \quad
C_{0101}= - \textstyle{\frac{1}{2}} C_{ijij}=\WD{ } .
\label{Wcomptsid}
\ee
The type D Weyl tensor is  thus completely determined by $\frac{m(m-1)}{2}$ independent components of  $\WDA{ij} $ and
$\frac{m^2(m^2-1)}{12}$ independent components of $C_{ijkl}$, where $n=m-2$.\footnote{\label{note_4D} In the standard $n=4$ (i.e., $m=2$) case these are essentially the imaginary and real part of $\Psi_2$. More specifically, {with the conventions of \cite{Stephanibook},} one has $\WDS{ij} =\frac{1}{2}\Phi\delta_{ij}$ with $\Phi=-2\mbox{Re}(\Psi_2)$, $\WDA{23} =\Phi_{23}=-\mbox{Im}(\Psi_2)$ as the only essential component of $\WDA{ij} $, while the $C_{ijks}$ reduce to the only non-trivial component $C_{2323}=-\Phi$.}

\section  {Static and stationary spacetimes }
\label{GID}

\subsection  {Static spacetimes}

Algebraically special spacetimes in higher dimensions are characterized by the existence
of preferred null directions - Weyl aligned null directions (WANDs).
A necessary and sufficient condition for a null vector $\bl$ being WAND in arbitrary
dimension is \cite{Milsonetal05,PraPra05}
\be
\ell^b \ell^c \ell_{[e}C_{a]bc[d}\ell_{f]}=0, \label{WANDeq}
\ee
where $C_{abcd}$ is the Weyl tensor.
Let us now assume that a spacetime of interest is algebraically special
and thus the equation    (\ref{WANDeq}) possesses a null solution $\bl= (\ell^t,\ell^A)$, $A=1\dots n-1$ (note that necessarily $\ell^t\neq0$ and at least one of the remaining components is also non-zero).

For static spacetimes the metric does not depend on the direction of time and
consequently the form of the metric and of the Weyl tensor remains unchanged
under the transformation $\tilde t = -t$.
Therefore, in these new coordinates equation  (\ref{WANDeq}) has the same form as in the original coordinates
and admits a {second} solution ${\tilde{\bn}} = ({{\ell^t}},\ell^A)$.
In the original coordinates
${{\bn}} = (-\ell^t,\ell^A)$.
Thus for static spacetimes the existence of a WAND $\bl$ implies the existence of a distinct WAND $\bn$
which in fact has the same order of alignment. The only Weyl types compatible with this property are types G, ${\rm I}_i$ and D {(or, trivially, O, i.e. conformally flat spacetimes)}. Therefore
\begin{proposition}
\label{propstatic}
All static spacetimes in arbitrary dimension are of Weyl types {\rm G}, ${\rm I}_i$ or {\rm D},
unless conformally flat.
\end{proposition}

In fact explicit examples of static spacetimes of
these Weyl types are known - charged static black ring (type G - \cite{OrtPra06}),
vacuum static black ring (type ${\rm I}_i$ - \cite{PraPra05}), the Schwarzschild-Tangherlini black hole
(type D - \cite{Pravdaetal04}) and the Einstein universe $\mathbb{R}\times S^{n-1}$ (type O - cf.~the results summarized in section~\ref{warped}). {Cf. also the static examples given in \cite{ColPel06}.}

Note that in four dimensions there is no  type G  and type I is equivalent to type ${\rm I}_i$ \cite{Coleyetal04,Milsonetal05}. Thus for $n=4$ only types I, D  and O are compatible
with static spacetimes. This was discussed already in \cite{petrov}
in the case of static, $n=4$, {vacuum } spacetimes (see also additional comments in \cite{Kinnersley:JMP69} {and in section~6.2 of \cite{Stephanibook}}).

\subsection {Stationary spacetimes }

One can  use  the same arguments  as above for stationary spacetimes with the metric remaining
unchanged under reflection symmetry involving time and some other coordinates.
 E.g. {in Boyer-Lindquist coordinates the} Kerr metric remains unchanged
under $\tilde{t}=-t$, $\tilde{\phi} = -{\phi}$ and  {$n=5$}
Myers-Perry under $\tilde{t}=-t$, $\tilde{\phi} = -{\phi}$,
$\tilde{\psi} = -{\psi}$ or, for general dimension,  Myers-Perry
under $\tilde{t}=-t$, $\tilde{\phi}_i = -{\phi}_i$. Note, however,
that in contrast to the static case, in some special stationary
cases one could in principle get from the original WAND $\bl$  a
``new'' WAND  ${\bn}= - {{\bl}}$ which represents the same null
direction. In order to deal with these special cases we note that
the {``divergence scalar'' (or, loosely speaking, ``expansion'',
since it does coincide with the standard expansion scalar in the
case of geodetic, affinely parameterized null directions)} of both
WANDs ${{\bn}}$ and ${{\bl}}$ related by reflection symmetry
is the same (as well as all the other optical scalars and the geodeticity parameters
- this also applies to the static case), i.e. $\ell^a_{\ ;a} = { n}^a_{\ ;a} $ while the
``expansion'' of  $- {{\bl}}$ is equal to $-\ell^a_{\ ;a}$.
Therefore for all ``expanding'' spacetimes ${{\bn}} \not= - \bl$.
Thus
\begin{proposition}
\label{propstationary}
In arbitrary dimension, all  stationary spacetimes with non-vanishing
divergence scalar (``expansion'') and
invariant under
appropriate reflection symmetry  are of Weyl types {\rm G}, ${\rm I}_i$ or {\rm D}, {unless conformally flat}.
\end{proposition}

Note also that it is shown in \cite{PraPraOrt07KS} that Kerr-Schild spacetimes {with the assumption $R_{00}=0$} are of type II (or more special) in arbitrary
dimension with the Kerr-Schild vector being the multiple WAND.
 Therefore all Kerr-Schild spacetimes that are either static or belong to the above mentioned class of stationary spacetimes are necessarily of type D. In particular, {\em the Myers-Perry metric in arbitrary dimension is thus of type~D}.\footnote{This was already known in the case $n=5$ \cite{Pravdaetal04,ColPel06}. Furthermore, it has been demonstrated recently in \cite{Hamamotoetal06} by explicit computation of the full curvature tensor that the family \cite{CheLuPop06} of higher dimensional rotating black holes with a cosmological constant and NUT parameter is of type D for any $n$. We observe in addition that, using the connection 1-forms given in \cite{Hamamotoetal06}, it is also straightforward to show (see the Appendix) that the mutiple WANDs (which are related by reflection symmetry) of all such solutions are twisting, expanding and shearing (except that the shear vanishes for $n=4$). The fact that the WANDs found in \cite{Hamamotoetal06} are {\em complex} is only due to the analytical continuation trick used in \cite{CheLuPop06} to cast the line element in a nicely symmetric form - the WANDs of the associated ``physical'' spacetimes are thus {\em real} after Wick-rotating back one of the coordinates.}

In addition to the rotating Myers-Perry black holes for $n \geq 4$, of
type D, we can mention a number of physically relevant solutions as explicit examples of spacetimes subject to Proposition~\ref{propstationary}.\footnote{It is straightforward to verify the ``reflexion symmetry'' of the metric we mention in this context. The ``expansion'' condition, instead, has not been verified explicitly in all cases. However, it is plausible that these spacetimes are indeed ``expanding'' since they contain as special limits or subcases solutions with expansion, e.g. Myers-Perry black holes (cf. section~\ref{subsec_MP}, \cite{Pravdaetal04} {and the preceding footnote}).} First,
 rotating vacuum black rings \cite{EmpRea02prl}, of type ${\rm I}_i$ \cite{PraPra05}. To our knowledge, no stationary
(non-static) type G solution has been so far
explicitly identified. It is, however, plausible to expect that a rotating charged black ring (so far unknown in the standard Einstein-Maxwell theory) will be of type G as its static counterparts. Further interesting examples
fulfilling our assumptions are expanding stationary
axisymmetric spacetimes with $n-2$ commuting Killing vector fields
\cite{Harmark04}, which contain, apart from the ($n=5$) black holes/rings mentioned above, also e.g. the recently obtained ``black saturn'' \cite{ElvangFiguerasSaturn},
doubly spinning black rings \cite{Pomeransky2006} and black di-rings \cite{IguchiMishima07}.  In any dimension also rotating uniform black strings/branes satisfy the assumptions of Proposition~\ref{propstationary} (see section~\ref{warped}), and so does the ansatz recently used in \cite{KleKunRad07} for the numerical construction of corresponding $n=6$ non-uniform solutions. Other examples are all the stationary solutions discussed in \cite{ColPel06} {and various black ring solutions reviewed in \cite{EmpRea06}}.

\subsection  {Remarks and ``limitations'' of the results}
\label{sub_sec_limitations}

First, it is worth observing that we have not used any field
equations for the gravitational field in the considerations
presented above and the results are thus purely geometrical.

Note that one can not relax the assumption $\ell^a_{\ ;a}\neq 0$
in the case of stationary spacetimes. For example, the special \pp
wave metric $\d s^2=g_{ij}\d x^i\d x^j-2\d u\d v-2H\d u^2$ such
that $H_{,u}=0$ (note that it is always $H_{,v}=0$ by the
definition of \pp waves) and $\pa_u\cdot\pa_u=-2H<0$ represents
stationary spacetimes {(cf., e.g., \cite{PodVes98czjp} for the
$n=4$ vacuum case) } that are invariant under reflection symmetry
($\tilde u=-u$, $\tilde v=-v$) and yet of type N
\cite{Coleyetal03}. In fact, the geodetic multiple WAND
$\bl=\pa_v$ is non-expanding (and $ {\bn}=-\bl$ is not a new
WAND).

Furthermore, if we assume a {\em null Killing vector field} $\bk$
instead of a timelike one we are led to different conclusions.
Namely, it is easy to show that $\bk$ must be geodetic, shearfree
and non-expanding, which for $R_{ab}k^a k^b=0$ implies that $\bk$
is a twistfree WAND \cite{OrtPraPra07}. We thus end up with a
subfamily of the Kundt class, for which (under the alignment
requirement $R_{ab}k^a\propto k_b$, obeyed e.g. in vacuum) the
algebraic type is II or more special \cite{OrtPraPra07} (cf.
section  ~24.4 of \cite{Stephanibook} for $n=4$). In particular, a
similar argument applies
locally at Killing horizons, where the type must thus be again II or
more special (provided $R_{ab}k^a\propto k_b$).\footnote{The proof is a bit more tricky in this case since the Killing vector is null only at the horizon. Still, one can adapt techniques used in \cite{LewPaw05,Hollandsetal06} for related investigations. Note that the horizon of higher dimensional
stationary black holes is indeed a Killing horizon (at least in the non-degenerate case) \cite{Hollandsetal06}.} This is in
agreement with the result of \cite{LewPaw05} for generic isolated
horizons. As an explicit example, vacuum black rings (which are of
type ${\rm I}_i$ in the stationary region) become locally of type
II on the horizon \cite{PraPra05}.

Finally, spacelike Killing vectors do not impose any
constraint on the algebraic type of the Weyl tensor, in general,
and all types are in fact possible. For example charged static
black rings are of type G, vacuum black rings of type I$_i$,
vacuum black holes of type D, and
they all admit at least one spacelike Killing vector; Kundt spacetimes can be constructed that admit axial symmetry with all types II, D, III and N being possible {(see, e.g., \cite{Stephanibook} for $n=4$)}.

\section{Direct/warped product spacetimes }
\label{warped}

{In this section we show that the algebraic types discussed above also characterize certain classes of direct/warped product geometries of physical relevance. In addition we discuss some optical properties of these spacetimes.}

\subsection  {Weyl tensor}

Let us consider two (pseudo-)Riemannian spaces $(M_1,g_{(1)})$ and $(M_2,g_{(2)})$ of dimension $n_1$ and $n_2$ ($n_1,n_2\ge1$ {and $n_1+n_2\ge 4$}), parameterized by coordinates $x^A$ ($A,B=0,\ldots,n_1-1$) and $x^I$ ($I,J=n_1,\ldots,n_1+n_2-1$), respectively. Using adapted coordinates $x^\mu$ ($\mu,\nu=0,\ldots,n_1+n_2-1$) constructed from the coordinates $x^A$ of $M_1$  and $x^I$ of $M_2$, we define the {\em direct product} $(M,g)$ to be the product manifold $M=M_1\times M_2$, of dimension $n=n_1+n_2$, equipped with the metric tensor $g(x^\mu)=g_{(1)}(x^A)\oplus g_{(2)}(x^I)$ defined (locally) by $g_{AB}=g_{(1)AB}$, $g_{IJ}=g_{(2)IJ}$, $g_{AI}=0$.
For the sake of definiteness, we shall assume hereafter that $(M_1,g_1)$ is Lorentzian and $(M_2,g_2)$ is Riemannian.

In general, any geometric quantity which can be split like the product metric (i.e., with no mixed components and with the $A[I]$ components depending only on the $x^A[x^I]$ coordinates) is called a ``product object'' (or ``decomposable''). Various interesting geometrical properties then follow \cite{Ficken39} and, in particular, the Riemann and Ricci tensors and the Ricci scalar are all decomposable.
As a consequence, {\em a product space is an Einstein space iff each factor is an Einstein space and their Ricci scalars satisfy $R_{(1)}/n_1=R_{(2)}/n_2$} \cite{Ficken39}.

Using the above coordinates it follows from the standard definition that the mixed components of the Weyl tensor are given by
\beqn
 \fl & & C_{ABCI}=C_{ABIJ}=C_{AIJK}=0 , \label{mixed1} \\
 \fl & & C_{AIBJ}=-\frac{1}{n-2}\left(g_{(1)AB}R_{(2)IJ}+g_{(2)IJ}R_{(1)AB}\right)+\frac{R_{(1)}+R_{(2)}}{(n-1)(n-2)}g_{(1)AB}g_{(2)IJ} , \label{mixed2}
\eeqn
where $R_{(1)AB}$ [$R_{(2)IJ}$] is the Ricci tensor of $(M_1,g_1)$ [$(M_2,g_2)$].
For the non-mixed components one has to distinguish the special cases $n_1=1,2$ (and the ``symmetric'' cases $n_2=1,2$, which we omit for brevity). If $n_1=1$ there are of course no non-mixed components $C_{ABCD}$ since now the $x^A$ span a one-dimensional space. If $n_1=2$ there is only one independent component, i.e. $C_{0101}$ (notice that here, exceptionally, 0 and 1 are not frame indices but refer to the coordinates $x^0$ and $x^1$ in the factor space $M_1$). For $n_1\ge 3$,
\beqn
 \fl   & & C_{ABCD}=C_{(1)ABCD}+\frac{2n_2}{(n-2)(n_1-2)}\left(g_{(1)A[C}R_{(1)D]B}-g_{(1)B[C}R_{(1)D]A}\right) \nonumber \label{nonmixed1} \\
\fl  & & \quad {}+\frac{2}{(n-1)(n-2)}\left[R_{(2)}-R_{(1)}\frac{n_2(n_2+2n_1-3)}{(n_1-1)(n_1-2)}\right]g_{(1)A[C}g_{(1)D]B} \qquad (n_1\ge 3) ,
\eeqn
where $C_{(1)ABCD}$ is the Weyl tensor of $(M_1,g_1)$, whereas the remaining non-mixed components are given for any $n_1\ge 1$ by
\beqn
 \fl & & C_{IJKL}=C_{(2)IJKL}+\frac{2n_1}{(n-2)(n_2-2)}\left(g_{(2)I[K}R_{(2)L]J}-g_{(2)J[K}R_{(2)L]I}\right) \nonumber \label{nonmixed2} \\
\fl  & & \quad {}+\frac{2}{(n-1)(n-2)}\left[R_{(1)}-R_{(2)}\frac{n_1(n_1+2n_2-3)}{(n_2-1)(n_2-2)}\right]g_{(2)I[K}g_{(2)L]J} \qquad (n_2\ge 3) ,
\eeqn
where $C_{(2)IJKL}$ is the Weyl tensor of $(M_2,g_2)$.
It is thus obvious that the Weyl tensor is not decomposable, in general. It turns out that {\em the Weyl tensor is decomposable iff both product spaces are Einstein spaces and $n_2(n_2-1)R_{(1)}+n_1(n_1-1)R_{(2)}=0$} (the latter condition is identically satisfied whenever $n_1=1$ or $n_2=1$, while for $n_1=2$ [$n_2=2$] it implies that $(M_1,g_1)$ [$(M_2,g_2)$] must be of constant curvature).
When the Weyl tensor is decomposable the only non-vanishing components take the simple form $C_{ABCD}=C_{(1)ABCD}$, $C_{IJKL}=C_{(2)IJKL}$.
Therefore, in particular, the product space is conformally flat iff both product spaces are of constant curvature and $n_2(n_2-1)R_{(1)}+n_1(n_1-1)R_{(2)}=0$.

Determining the possible algebraic types of the Weyl tensor
requires considering various possible choices for the dimension
$n_1$ of the Lorentzian factor.

If $n_1=1$, the full metric can
always be cast in the special static form $\d s^2=-\d t^2+g_{IJ}\d
x^I\d x^J$. Recalling the result of section~\ref{GID}, the Weyl
tensor can thus only be of type G, I$_i$, D or O. In particular,
one can show that $C_{0i1j}=C_{0j1i}$, so that for direct product  spacetimes with
$n_1=1$ one has $\WDA{ij} =0$ identically.

If $n_1\ge2$, it is convenient to adapt the null frame~(\ref{frame}) to the natural product structure, so that $g_{a b} = 2\ell_{(a}n_{b)} + \delta_{\hat A \hat B} m^{(\hat A)}_a m^{(\hat B)}_b+\delta_{\hat I\hat J} m^{(\hat I)}_a m^{(\hat J)}_b$ (where $\hat A, \hat B=2,\ldots,n_1-1$, $\hat I, \hat J=n_1,\ldots,n-1$ are now frame indices, and the frame vectors do not have mixed coordinate components, e.g. $\ell^I=0=n^I$ etc.). From~(\ref{nonmixed1}) and (\ref{nonmixed2}) it thus follows that $C_{ABCD}$ and $C_{IJKL}$ do not give rise to mixed frame components, and from~(\ref{mixed2}) that $C_{AIBJ}$ does not give rise to non-mixed frame components. Hence the only non-vanishing mixed components are (ordered by boost weight)
\beqn
 \fl & & C_{0\hat I 0\hat J}=-\frac{1}{n-2}R_{(1)00}\delta_{\hat I\hat J}, \qquad C_{0\hat I \hat A\hat J}=-\frac{1}{n-2}R_{(1)0\hat A}\delta_{\hat I\hat J}, \nonumber \\
 \fl & & C_{0\hat I 1\hat J}=-\frac{1}{n-2}\left(R_{(2)\hat I\hat J}+R_{(1)01}\delta_{\hat I\hat J}\right)+\frac{R_{(1)}+R_{(2)}}{(n-1)(n-2)}\delta_{\hat I\hat J}, \nonumber \label{mixed_frame} \\
 \fl & & C_{\hat A\hat I \hat B\hat J}=-\frac{1}{n-2}\left(R_{(2)\hat I\hat J}\delta_{\hat A\hat B}+R_{(1)\hat A\hat B}\delta_{\hat I\hat J}\right)+\frac{R_{(1)}+R_{(2)}}{(n-1)(n-2)}\delta_{\hat A\hat B}\delta_{\hat I\hat J}, \\
 \fl & & C_{1\hat I \hat A\hat J}=-\frac{1}{n-2}R_{(1)1\hat A}\delta_{\hat I\hat J}, \qquad C_{1\hat I 1\hat J}=-\frac{1}{n-2}R_{(1)11}\delta_{\hat I\hat J}.  \nonumber
\eeqn

The non-mixed frame components are given for $n_1=2$ by
\be
 \fl C_{0101}=-\frac{1}{2(n_2+1)}\left[(n_2-1)R_{(1)}+\frac{2R_{(2)}}{n_2}\right] \qquad (n_1=2) \label{n1=2},
\ee
and for $n_1\ge 3$ by
\beqn
 \fl & & C_{0\hat A 0\hat B}=C_{(1)0\hat A 0\hat B}+\frac{n_2}{(n-2)(n_1-2)}
        R_{(1)00}\delta_{\hat A\hat B}, \nonumber \label{nonmixed_frame} \\
 \fl & & C_{010\hat A}=C_{(1)010\hat A}-\frac{n_2}{(n-2)(n_1-2)}
        R_{(1)0\hat A} , \nonumber \\
 \fl & & C_{0\hat A \hat B\hat C}=C_{(1)0\hat A \hat B\hat C}-\frac{2n_2}{(n-2)(n_1-2)}
        R_{(1)0[\hat C}\delta_{\hat B]\hat A}, \nonumber \\
 \fl & & C_{0101}=C_{(1)0101}-\frac{2n_2}{(n-2)(n_1-2)}R_{(1)01} \nonumber \\
        \fl  & & \qquad \qquad
        {}-\frac{1}{(n-1)(n-2)}\left[R_{(2)}-R_{(1)}\frac{n_2(n_2+2n_1-3)}{(n_1-1)(n_1-2)}\right]       , \nonumber \\
 \fl & & C_{0 1\hat A\hat B}=C_{(1)0 1\hat A\hat B} \hspace{6.8cm} (n_1\ge 3) , \\
 \fl & & C_{0\hat A 1 \hat B}=C_{(1)0\hat A 1 \hat
        B}+\frac{n_2}{(n-2)(n_1-2)}\left(R_{(1)\hat A\hat B}+R_{(1)01}\delta_{\hat A\hat
        B}\right) \nonumber \\
        \fl  & & \qquad \qquad {}+\frac{1}{(n-1)(n-2)}\left[R_{(2)}-R_{(1)}\frac{n_2(n_2+2n_1-3)}{(n_1-1)(n_1-2)}\right]\delta_{\hat A\hat B} , \nonumber \\
 \fl & & C_{\hat A\hat B \hat C\hat D}=C_{(1)\hat A\hat B \hat C\hat
        D}+\frac{2n_2}{(n-2)(n_1-2)}\left(R_{(1)\hat B[\hat D}\delta_{\hat C]\hat A}-R_{(1)\hat             A[\hat D}\delta_{\hat C]\hat B}\right) \nonumber \\
\fl  & & \qquad \qquad  {}+\frac{2}{(n-1)(n-2)}\left[R_{(2)}-R_{(1)}\frac{n_2(n_2+2n_1-3)}{(n_1-1)(n_1-2)}\right]\delta_{\hat B[\hat D}\delta_{\hat C]\hat A} , \nonumber \\
\fl & & C_{\hat I\hat J\hat K\hat L}=C_{(2)\hat I\hat J\hat K\hat
            L}+\frac{2n_1}{(n-2)(n_2-2)}\left(\delta_{\hat I[\hat K}R_{(2)\hat L]\hat
            J}-\delta_{\hat J[\hat K}R_{(2)\hat L]\hat I}\right) \nonumber \\
            \fl  & & \qquad \qquad
{}+\frac{2}{(n-1)(n-2)}\left[R_{(1)}-R_{(2)}\frac{n_1(n_1+2n_2-3)}{(n_2-1)(n_2-2)}\right]\delta_{\hat I[\hat K}\delta_{\hat L]\hat J}  , \nonumber \\
 \fl & & C_{1\hat A \hat B\hat C}=C_{(1)1\hat A \hat B\hat C}-\frac{2n_2}{(n-2)(n_1-2)}
        R_{(1)1[\hat C}\delta_{\hat B]\hat A}, \nonumber \\
 \fl & & C_{101\hat A}=C_{(1)101\hat A}-\frac{n_2}{(n-2)(n_1-2)}
        R_{(1)1\hat A} , \nonumber \\
 \fl & & C_{1\hat A 1\hat B}=C_{(1)1\hat A 1\hat B}+\frac{n_2}{(n-2)(n_1-2)}
R_{(1)11}\delta_{\hat A\hat B}. \nonumber
\eeqn
(The expression for $C_{\hat I\hat J\hat K\hat L}$ holds only when $n_2\ge 3$, while for $n_2=2$ one gets only one component $C_{2323}$ similar to~(\ref{n1=2}).)

For $n_1=2$ the Weyl tensor of $(M_1,g_1)$ of course vanishes, and in addition we have $R_{(1)00}=0=R_{(1)11}$ identically (any 2-space satisfies $2R_{(1)AB}=R_{(1)}g_{(1)AB}$). Therefore among the above components (\ref{mixed_frame}) and (\ref{n1=2}) only the boost weight zero components
$C_{0\hat I 1\hat J}$ and $C_{0101}$ survive, so that the corresponding
spacetime can be only of type D (or conformally flat), and
 $\bl$ and $\bn$, as chosen above, are multiple WANDs. Note
also that $\WD{ij} $ reduces to $\WD{\hat I\hat J} =C_{0\hat I
1\hat J}=C_{0\hat J 1\hat I}$ in this case, therefore $\WDA{ij}
=0$. As an example, the higher dimensional electric Bertotti-Robinson solutions
fall in this class, cf., e.g, \cite{FreRub80,CarDiaLem04}.

For $n_1=3$, again the Weyl tensor of $(M_1,g_1)$ vanishes. With the additional assumption that $(M_1,g_1)$ is Einstein, we get  $R_{(1)00}=R_{(1)11}=R_{(1)0\hat A}=R_{(1)1\hat A}=0$ (here $\hat A=2$ only), and as above the Weyl tensor is of type D with $\WDA{ij} =0$.

Similarly, for any $n_1>3$, if $(M_1,g_1)$ is an Einstein space
the only non-zero mixed Weyl components (\ref{mixed_frame}) will
have boost weight zero, and the non-mixed components (\ref{nonmixed_frame}) simplify considerably. As a particular consequence, if
$(M_1,g_1)$ is an Einstein space of type D, $(M,g)$ will also be
of type D ({but now $\WDA{ij} \neq0$, in general}) -
this is the case, for example, of uniform black strings/branes
(either static or rotating, see also the discussion concluding
this section). If $(M_1,g_1)$ is of constant curvature, $(M,g)$
will be of type D with $\WDA{ij} =0$ (or O) - this
includes the higher dimensional magnetic Bertotti-Robinson
solutions \cite{FreRub80}. One can consider other special cases
using similar simple arguments.

A spacetime conformal to a { direct product} spacetime is called a {\em warped  product}
spacetime if the conformal factor depends only on one of the two coordinate sets
$x^A$, $x^I$ (see e.g. \cite{Stephanibook}). Obviously, the algebraic type of two conformal
spaces is the same.\footnote{This is true also for doubly warped  product spacetimes discussed in \cite{RamosVazJMP03}, so that Propositions \ref{prop_warped1} and \ref{prop_warped2} hold also in that case.} Some of the results presented above can thus be straightforwardly generalized to the more general case of warped products. For example,

\begin{proposition}
\label{prop_warped1}
In arbitrary dimension, a warped spacetime with a one-dimensional Lorentzian (timelike) factor can be only of type G, I$_i$, D (with $\WDA{ij} =0$) or O.
\end{proposition}

This case includes, in particular, the conclusion of
section~\ref{GID} for static spacetimes. As warped
non-static/non-stationary examples we can mention the de~Sitter
universe (in global coordinates) and FRW cosmologies.
For $n=4$
Proposition~\ref{prop_warped1} reduces to a result of
\cite{CardaC93}.

Furthermore,
\begin{proposition}
\label{prop_warped2}
 In arbitrary dimension, a warped spacetime with a two-dimensional Lorentzian factor can be only of type D (with $\WDA{ij} =0$) or O.
\end{proposition}

Cf.~again \cite{CardaC93} for $n=4$. Notice that in this case the
line element can always be cast in one of the two (conformally
related) forms $\d s^2=2A(u,v)\d u\d v+f(u,v)h_{IJ}(x)\d x^I\d
x^J$ or $\d s^2=2\tilde f(x)A(u,v)\d u\d v+g_{IJ}(x)\d x^I\d x^J$
(so that multiple WANDs are given by $\pa_u$ and $\pa_v$), which
include a number of known spacetimes. For example, the first
possibility includes all spherically symmetric spacetimes,
hence as a special case of Proposition~\ref{prop_warped2} we have
\begin{proposition}
\label{prop_spherical}
In arbitrary dimension, a spherically symmetric spacetime is of type D (with $\WDA{ij} =0$) or O.
\end{proposition}

For $n=4$ this has been known for a long time (see e.g.
\cite{PleSta68} and sections~15.2, 15.3 of
\cite{Stephanibook}), and in this case $\WDA{ij} =0$ means that
$\Psi_2$ is real (see the footnote on p.~\pageref{note_4D}). For $n>4$ this result has been proven in
\cite{HorRos97} in the static case.

Other properties of decomposable Weyl tensors were discussed in \cite{Coleyetal04}.

\subsection  {``Factorized'' geodetic null vector fields}

Let us define an $n$-dimensional spacetime $(M,g)$ as the warped
product of an $n_1$-dimensional Lorentzian space $(M_1,g_{(1)})$
and an $n_2$-dimensional Riemannian space  $(M_2,g_{(2)})$, with
$n=n_1+n_2$ as in the preceding subsection. Hereafter we shall
assume $n_1\ge 2$. Using the adapted coordinates defined above,
the metric can take one of the following two forms \beqn
 \d s^2=g_{AB}\d x^A\d x^B+f(x^A)h_{IJ}\d x^I\d x^J , \label{warped1} \\
 \d s^2=\tilde f(x^I)h_{AB}\d x^A\d x^B+g_{IJ}\d x^I\d x^J \label{warped2} ,
\eeqn where $g_{AB},h_{AB}=g_{(1)AB}$ depend only on the $x^A$
coordinates and $g_{IJ},h_{IJ}=g_{(2)IJ}$ only on the $x^I$
coordinates.

Given a null vector $\bl_{(1)}=\ell_{(1)}^A\pa_A$ of $M_1$, this can
be ``lifted'' to define a null vector $\bl$ of $M$ with covariant
components $\ell_A=\ell_{(1)A}$ (functions of the $x^A$ only) and
$\ell_I=0$. From equations (\ref{warped1}), (\ref{warped2}) it follows
that if $\bl_{(1)}$ is geodetic (and affinely parameterized) in
$M_1$ then $\bl$ is automatically geodetic (and affinely
parameterized) in $M$. We can thus ``compare'' the
optical scalars of $\bl_{(1)}$ in $M_1$ with those of $\bl$ in
$M$.  For the warped metric~(\ref{warped1}), with the
definitions~(\ref{scalars}) one finds \beqn
 & & \sigma^2=\sigma_{(1)}^2+\frac{(n_1-2)n_2}{n_1+n_2-2}\left[\theta_{(1)}-\frac{1}{2}(\ln
 f)_{,A}\ell^A\right]^2 , \nonumber \\
 & & \theta=\frac{1}{n_1+n_2-2}\left[(n_1-2)\theta_{(1)}+\frac{n_2}{2}(\ln
 f)_{,A}\ell^A\right] , \label{scalars_w1} \\
& & \omega^2=\omega_{(1)}^2 , \nonumber \eeqn where
$\sigma_{(1)}^2$, $\theta_{(1)}$ and  $\omega_{(1)}^2$ are the
optical scalars of $\bl_{(1)}$ in $(M_1,g_{(1)})$. For the warped
metric~(\ref{warped2}) one has
 \beqn
 & & \sigma^2=\tilde f^{-2}\left[\sigma_{(1)}^2+\frac{(n_1-2)n_2}{n_1+n_2-2}\theta_{(1)}^2\right] , \nonumber \\
 & & \theta=\frac{n_1-2}{n_1+n_2-2}\tilde f^{-1}\theta_{(1)} , \label{scalars_w2} \\
& & \omega^2=\tilde f^{-2}\omega_{(1)}^2 . \nonumber  \eeqn

The special case of direct products is recovered for $f, \tilde
f=\mbox{const.}$ (which can be rescaled to 1), in which case the
shear of the full spacetime originates in the shear and expansion
of the Lorentzian factor (while expansion and twist are
essentially the same as in $(M_1,g_{(1)})$).

Note that for $n_1=2$ the definitions~(\ref{scalars}) for
$\sigma_{(1)}^2$ and $\theta_{(1)}$ become formally singular
because of the normalization, but for a Lorentzian 2-space (e.g.,
$\d s^2=2A(u,v)\d u\d v$ with the geodetic null vector
$\bl=A^{-1}\pa_v$) one has
$\ell_{(a;b)}\ell^{(a;b)}=\ell^a_{\;;a}=\ell_{[a;b]}\ell^{a;b}=0$,
so that we can essentially take
$\sigma_{(1)}^2=\theta_{(1)}=\omega_{(1)}=0$ and
formulae~(\ref{scalars_w1}), (\ref{scalars_w2}) still hold.

The results of this section   can be applied to several known
solutions. For example, static [rotating] black strings and branes
(i.e, direct products of Schwarzschild [Kerr] cross a flat space)
are type D vacuum spacetimes with two shearing, expanding,
twistfree [twisting] multiple WANDs. As such, they clearly
``violate'' the Golberg-Sachs theorem. In addition, spherically
symmetric solutions in any dimensions (which necessarily take the metric
form~(\ref{warped1}) with $n_1=2$) are type D spacetimes with two
shearfree, expanding, twistfree multiple WANDs (independently of
any specific field equations; in the ``exceptional case'' $(\ln
 f)_{,A}\ell^A=0$ the vector $\bl$ is non-expanding, e.g. for Bertotti-Robinson/Nariai geometries, or for null generators of horizons).

\section  {Type D {Einstein} spacetimes in higher dimensions}
\label{DHD}

From the results of the previous sections it follows that type D spacetimes are the simplest
non-trivial examples of static/stationary (``expanding'' and with an appropriate reflection)/warped spacetimes.
Therefore we will focus on type D spacetimes in general (without assuming staticity etc.).
Recall
that the quantities/symbols used below (e.g. $\WD{ij} $, $L_{ij}$, $D$) are defined in section \ref{prelim}.

\subsection  {Algebraic conditions following from the Bianchi equations}

Various contractions of Bianchi identities
\be
R_{abcd;e} + R_{abde;c} + R_{abec;d} = 0 \label{Bia}
\ee
lead to a set of first-order PDEs for {frame} components of the {Riemann} tensor given in Appendix B of \cite{Pravdaetal04}.
{In the following we shall concentrate on Einstein spaces (defined by $R_{ab}=\frac{R}{n}g_{ab}$), for which the same set of equations holds unchanged also for components of the Weyl tensor.} In case of
algebraically special spacetimes,
some of these differential equations reduce to algebraical equations due to the vanishing of some components of the Weyl tensor.
Here we derive algebraic conditions following from the Bianchi equations for type D {Einstein} spacetimes.
 These conditions
will be employed in subsequent sections.

In particular, by contracting (\ref{Bia}) with $\bm{i} $, $\bl$,
$\bm{j} $, $\bm{k} $ and $\bl$ (equation  (B.8) in  \cite{Pravdaetal04})
and assuming to have a type D Einstein space we get the
first algebraic condition
\be
\WD{ij} L_k - \WD{ik} L_j + 2 \WDA{kj} L_i - C_{isjk} L_s = 0,
\label{Bia8}
\ee
where we denoted $L_{i0}$ by $L_i$. We will also denote  $L_i L_i$ by $L$.

The second algebraic equation   follows from equation  (B.15,\cite{Pravdaetal04})
\bea
\fl
0 = 2 \left( \WDA{jk} L_{im} + \WDA{mj} L_{ik} + \WDA{km} L_{ij}   + \WD{ij} A_{mk}
  + \WD{ik} A_{jm} + \WD{im} A_{kj} \right) \nonumber \\
 {}+ C_{isjk} L_{sm} + C_{ismj} L_{sk} + C_{iskm} L_{sj}  \label{algcomb15}
\eea
and contraction  of $k$ with $i$ leads to
\bea
 0&=&S \WDA{mj} +\WD{ } A_{jm}-(\WDS{mi} + \WDA{mi} ) S_{ij}
+ (\WDS{ji} + \WDA{ji} ) S_{im}
\nonumber\\
&&{}+2(\WDA{im} A_{ij}
-\WDA{ij} A_{im})
+\textstyle{\frac{1}{2}}C_{ismj}A_{si}. \label{eqB15i=k}
\eea

By contracting $m$ with $j$ in equation  (B.12) from \cite{Pravdaetal04}
we get
\bea
2 D \WDS{ik} &=& 4 \WDA{ij} A_{kj} + \WD{kj} L_{ij} +
\WD{ji} L_{jk} - \WD{ki} S  -  \WD{ } L_{ik} - 2 \WDS{is} L_{sk}
\nonumber \\ && {}- 2 \WDS{sk} \Ms_{i0} - 2 \WDS{is} \Ms_{k0} +
C_{ijks} L_{sj}, \label{DPhi1}
\eea
where we employed  $C_{iskj}
\Ms_{j0} + C_{ijks} \Ms_{j0} = 0$ ($\M{s}{j}{0}+\M{j}{s}{0}=0$, cf.~\cite{Pravdaetal04}).

The symmetric part of equation  (B.5,\cite{Pravdaetal04}) and equation (B.3) (that is equivalent to the  antisymmetric part
of (B.5)) give, respectively,
\be
\fl
2 D \WDS{ik} = - 2 \WD{} S_{ik} + (-2 \WD{is} + \WD{si} ) L_{sk}
+ (-2 \WD{ks} + \WD{sk} ) L_{si} %\nonumber \\
- 2 \WDS{sk}  \Ms_{i0} - 2 \WDS{is}  \Ms_{k0}\! ,\label{DPhi2}
\ee
\be
\fl
2 D \WDA{ik} = - 2 \WD{} A_{ik} + (-2 \WD{is} + \WD{si} ) L_{sk}
- (-2 \WD{ks} + \WD{sk} ) L_{si} %\nonumber \\
- 2 \WDA{sk}  \Ms_{i0} + 2 \WDA{si}  \Ms_{k0}\! .\label{DPhi2_as}
\ee

By subtracting (\ref{DPhi2}) from (\ref{DPhi1}) we finally obtain
the third algebraic equation   \bea \fl 0 =   -\WD{ki} S + \WD{}
L_{ki} + \WD{kj} L_{ij} + 4 \WDA{ij} A_{kj} + (2\WD{kj} - \WD{jk}
) L_{ji}
 + 2 \WDA{ij} L_{jk} + C_{ijks} L_{sj} . \label{algcomb}
\eea
Its antisymmetric part is, thanks to $C_{ikjm}A_{mj}=2C_{ijks}A_{sj}$, equal to equation  (\ref{eqB15i=k}) and
its symmetric part reads
\be
\fl
0=-S \WDS{ik} + \WD{ } S_{ik} +\WDS{ij} S_{jk} +\WDS{kj} S_{ij} +3(\WDA{ij} S_{jk} +\WDA{kj} S_{ji})
+C_{ijks} S_{sj} .\label{eqB12B5sym}
\ee

Equations (\ref{Bia8}), (\ref{eqB15i=k}) and (\ref{eqB12B5sym}) will be extensively used in the following
sections.

{In passing, let us observe here in what sense the $n=4$ case is unique. Recalling the footnote on p.~\pageref{note_4D}, from (\ref{Bia8}) we get $L_i=0$ (geodetic property) unless $\Phi_{ij}=0$ (trivial case of zero Weyl tensor); equation   (\ref{eqB15i=k}) is identically satisfied (noting that necessarily $\WDA{ij} \propto A_{ij}$ when $n=4$); equation   (\ref{eqB12B5sym}) implies $S_{ij}\propto\delta_{ij}$ (vanishing shear) again unless $\Phi_{ij}=0$. Thus for $n=4$ we correctly recover the standard Goldberg-Sachs result (here restricted to type D spacetimes) that multiple WANDs (PNDs) are geodetic and shearfree in vacuum (and Einstein) spaces \cite{Stephanibook}}.
The situation in higher dimensions, which is qualitatively different from the $n=4$ case, is studied in the following sections.

\subsection  {WANDs in ``generic'' vacuum type D and II spacetimes in arbitrary dimension are geodetic}

In this section   we study equation   (\ref{Bia8}) in order to determine under which circumstances the multiple WAND $\bl$ is geodetic.

By contracting $i$ with $k$ in (\ref{Bia8}) and using (\ref{Wcomptsid}) we get
\be
\left( 3 \WDA{ij} - \WDS{ij}  \right) L_i = \WD{} L_j
\label{Bia8a1}
\ee
and after multiplying (\ref{Bia8a1}) by $L_j$ we obtain
\be
\WDS{ij} L_i L_j = - \WD{} L.
\label{Bia8a2}
\ee
By multiplying  (\ref{Bia8}) by $L_i L_j$ and using (\ref{Bia8a2}) we get
\be
L \left( 3 \WDA{ik} L_i + \WDS{ik} L_i + \WD{} L_k \right) = 0.
\ee
Thus either $L=0$ or
\be
\left( - 3 \WDA{ij} - \WDS{ij}  \right) L_i = \WD{} L_j .
\label{Bia8a3} \ee By adding and subtracting  (\ref{Bia8a1}) and
(\ref{Bia8a3}) we get
\be
\WDS{ij} L_i = - \WD{} L_j, \qquad \WDA{ij} L_i =0. \label{Bia8a4}
\ee
Finally multiplying (\ref{Bia8}) by
$L_i$ and using (\ref{Bia8a4}) we get
\be
L \WDA{ij} = 0.\label{LiPhiA}
\ee
This implies that for a   type D vacuum spacetime  with non-vanishing $\WDA{ij} $
in arbitrary dimension corresponding WANDs are geodetic.

In the case with  vanishing $\WDA{ij} $, let us choose a frame in which $\WDS{ij} $ is diagonal
$\WDS{ij} =\mbox{diag} \{ p_{(2)},\ p_{(3)},\ \dots,\  p_{(n-1)}\}$. Then from the first equation   (\ref{Bia8a4}) it follows
\be
(p_{(i)}+\WD{} )L_i =0,
\ee
where (from now on) we do {\it not} sum over indices in brackets. If  $p_{(i)}\not =-\WD{} $, $\forall i$, then  $L_i =0$, $\forall i$, i.e. $\bl$ is geodetic.

Note that so far we have employed only equation   (\ref{Bia8}), which corresponds to equation  (B.8) in \cite{Pravdaetal04} and which does not contain Weyl tensor components with negative boost order.
Consequently, the same conclusions hold also for type II Einstein spacetimes.

\begin{proposition}
\label{propgeodetic}
 In arbitrary dimension, multiple WANDs  of  type II and D  Einstein spacetimes are geodetic
 if at least one of the following conditions is satisfied:\\
 i) $\WDA{ij} $ is non-vanishing; \\
 ii) for all eigenvalues of $\WDS{ij} $: $p_{(i)}\not= -\WD{} $.
\end{proposition}

{Note that the above argument can not be extended to more special algebraic classes of spacetimes since it relies on the fact that some Weyl components with boost weight zero are non-vanishing. However,} it was already shown in \cite{Pravdaetal04} that multiple WANDs in
type N and III vacuum spacetimes are geodetic  {(in that case with
no need of extra assumptions)}. Therefore we can conclude that
{under most} ``generic'' conditions  multiple WANDs are  geodetic.
Note, however, that {certain} special type-D {vacuum solutions} with $\WDA{ij}
= 0$ and $p_{(i)}=-\WD{} $ (for some $i$)  admit non-geodetic multiple WANDs. Explicit example of such
spacetime is given in section   \ref{nongeod}.

\subsection  {Vacuum type D spacetimes with a ``shearfree'' WAND}

\label{shearfreeD}

The  algebraic equations (\ref{eqB15i=k}) and (\ref{eqB12B5sym}) are
quite complicated in general dimension and thus here we will limit ourselves to the ``shearfree'' case. {This is of interest since it includes, for instance,  the Robinson-Trautman solutions containing static black holes \cite{PodOrt06}}.

With the ``shearfree'' condition
\be
S_{ij}=\textstyle{\frac{S}{n-2}} \delta_{ij}, \label{shearfree}
\ee
equation (\ref{eqB12B5sym}) leads for $S\neq 0$ to
\be
\WDS{ij} =\textstyle{\frac{\WD{ } }{n-2}}\delta_{ij} \qquad (S\neq 0), \label{Phidelta}
\ee
{whereas it is identically satisfied for $S=0$. In the rest of this subsection we thus consider only
the ``expanding'' case $S\neq 0$. }
For $\WDS{ij} $ in the form (\ref{Phidelta}) with $\WD{} \not= 0$ the condition ii) of Proposition \ref{propgeodetic}
is satisfied and thus
the WAND $\bl$ is geodetic.
\begin{proposition}
\label{propshearfree}
In arbitrary dimension, multiple ``shearfree'' and ``expanding'' WAND in a type D Einstein spacetime is geodetic whenever $\WD{ij} \not= 0$.
\end{proposition}

Note that  $\WD{ij} $ has to be non-zero for type D spacetimes in four and five
dimensions. Thus all such shearfree WANDs are geodetic.\footnote{In fact, for $n=4$ from the Goldberg-Sachs theorem we already knew that all multiple WANDs are automatically shearfree and geodetic.} On the other hand,
spacetimes with  $\WD{ij} = 0$ are not necessarily conformally flat
for $n>5$ ($C_{ijkl}$ can be non-vanishing, and in that case equation (\ref{Bia8}) reduces to $C_{isjk} L_s = 0$) and in fact in section
\ref{nongeod} we will present an example of such type D vacuum spacetime with a non-geodetic  {``shearfree''} multiple WAND.

Furthermore, using (\ref{shearfree}) and (\ref{Phidelta}), equation   (\ref{eqB15i=k}) reads
\be
0=\textstyle{\frac{n-4 }{n-2}} S\WDA{ij} +\WD{ } A_{ji} +2(\WDA{ki} A_{kj} +\WDA{jk} A_{ki}  )
+\textstyle{\frac{1 }{2}} C_{kmij}A_{mk}.
\ee
{As mentioned above this is identically satisfied for $n=4$. For $n>4$ (and $S\neq 0$), if one assumes $A_{ij}=0$ it gives $\WDA{ij} =0$, while assuming $\WDA{ij} =0$ leads to $C_{kmij}A_{mk} =2\WD{ } A_{ij}$. On the other hand, from equation   (\ref{DPhi2_as}) with (\ref{shearfree}) and (\ref{Phidelta}) we see that $\WDA{ij} =0$ implies $A_{ij}=0$, unless $\Phi=0$ (in which case the full $\Phi_{ij}$ would be zero). We can thus summarize these results in}
\begin{proposition}
\label{propshearfree2}
For a multiple ``shearfree'' and ``expanding'' WAND in a type D Einstein spacetime in $n>4$ dimensions the following implications hold
\begin{enumerate}
 \item $A_{ij}=0 \quad  \Rightarrow  \quad \WDA{ij} =0 . \label{sfA}$
 \item $\WDA{ij} =0, \  \WDS{ij} \neq 0 \quad \Rightarrow  \quad  A_{ij}=0.$
 \item $\WDA{ij} =0, \  \WDS{ij} = 0 \quad \Rightarrow  C_{kmij}A_{mk} =0.$
\end{enumerate}
\end{proposition}

Note that for an arbitrary odd-dimensional spacetime with a geodetic and shearfree WAND one has $A_{ij}=0$ \cite{OrtPraPra07} and thus in the expanding case, $\theta\neq 0$, by~(\ref{sfA}) $\WDA{ij} $ also necessarily vanish.
{Note also that the assumptions of~(\ref{sfA})} (i.e., $\sigma_{ij}=0=A_{ij}$, $\theta\neq 0$) uniquely identify the Robinson-Trautman spacetimes (which are of type D for $n>4$) {in any dimensions} and indeed $\WDA{ij} =0$ for the corresponding Weyl tensor \cite{PodOrt06}.
In general $\WDS{ij} =\frac{\WD{ } }{n-2}\delta_{ij}\neq0$ for Robinson-Trautman solutions \cite{PodOrt06} and by Proposition~\ref{propshearfree} the multiple WANDs are thus geodetic, however, in the next subsection    we present a very special Robinson-Trautman solution with vanishing $\WDS{ij} $ and with a non-geodetic WAND.

\subsection  {An example of type D vacuum spacetimes with a non-geodetic WAND }
\label{nongeod}

The conclusions in the preceding subsections about the geodetic
character of multiple WANDs can not be ({in contrast to the $n=4$
case}) extended to the most general case. In fact, here we point out that a
special subclass of the Robinson-Trautman solutions
\cite{PodOrt06} in $n\ge 7$ dimensions represents type D vacuum
spacetimes (with a possible cosmological constant) for which one
of the multiple WANDs is non-geodetic.
Namely, let us consider the
vacuum family \cite{PodOrt06,Ortaggio07} \beqn
 & & \d s^2=r^2h_{ij}\d x^i\d x^j-2\d u\d r-2H\d u^2 , \nonumber \label{RT_metric} \\
 & & 2H=K-2r(\ln P)_{,u}-\frac{2\Lambda}{(n-2)(n-1)}\,r^2   \qquad (K=0, \pm 1) ,
\eeqn
where $P^2=(\det h_{ij})^{1/(2-n)}$ and $h_{ij}$ represents an arbitrary $(n-2)$-dimensional Einstein space ($i,j=2\ldots,n-1$ are, exceptionally, coordinate indices in this subsection). Using a suitable frame based on the null vectors
\be
 \bl=\pa_r , \qquad \bn=-\pa_u+H\pa_r ,
 \label{null_vec}
\ee
the only non-vanishing components of the Weyl tensor have boost weight zero and are given by \cite{PodOrt06}
\be
 C_{ijkl}=r^2({\cal R}_{ijkl}-2K h_{i[k}h_{l]j}) ,
\label{Weylvac2}
\ee
where ${\cal R}_{ijkl}$ is the Riemann tensor associated to $h_{ij}$. This implies that the spacetime~(\ref{RT_metric}) is of type D, with $\WD{ij} =0$, and that both $\bl$ and $\bn$ are multiple WANDs. Now, the vector $\bl$ is geodetic, shearfree and twistfree by construction \cite{PodOrt06}. Next, one can easily show that
\be
 \nabla_{\mbox{\scriptsize$\bn$}}\bn=-H_{,r}\bn+H_{,i}\d x^i,
\ee where, by~(\ref{RT_metric}), $H_{,i}=-r(\ln P)_{,ui}$.
Therefore $\bn$ is geodetic if and only if $(\ln
P)_{,ui}=0\Leftrightarrow P=p_1(u)p_2(x^2,x^3,\ldots)$. For a
general (non-factorized) function $P$ the multiple WAND $\bn$ is
thus non-geodetic (one can also easily check that it ``shearfree'', ``twistfree'' and ``expanding''). A simple
explicit example of such spacetimes is obtained by extending to
any $n\ge 7$ the $n=7$ dimensional solution discussed in
\cite{Ortaggio07}, i.e. by taking in eq.~(\ref{RT_metric}) \beqn
 \fl K=-1 , \qquad P=f(u,z)^{-1/2}\left[\rho^{n-5}(\det\eta_{\alpha\beta})^{1/2}\right]^{1/(2-n)} , \nonumber \\
 \fl h_{ij}\d x^i\d x^j=f(u,z)\Bigg[\d z^2+V(\rho)\d\tau^2+\frac{1}{V(\rho)}\d\rho^2+\rho^2\eta_{\alpha\beta}\d x^\alpha\d x^\beta\Bigg], \label{einstein_h} \\
 \fl f(u,z)=\frac{4b(u)e^{2z/l}}{l^2[e^{2z/l}-b(u)]^2} , \qquad V(\rho)=\left(1-\frac{\mu}{\rho^{n-6}}-\frac{\rho^2}{l^2}\right)\label{einstein_f}, \nonumber
\eeqn
where $z\equiv x^2$, $\tau\equiv x^3$, $\rho\equiv x^4$, $\eta_{\alpha\beta}=\eta_{\alpha\beta}(x^5,x^6,\ldots)$ is the metric of an $(n-5)$-dimensional unit sphere ($\alpha,\beta=5,\ldots,n-1$),
 $\mu$ and $l$ are constants and $b(u)>0$ is an arbitrary function. The multiple WAND $\bn$ is non-geodetic as long as $\d b/\d u\neq0$. Note that there is not contradiction with the results of the previous subsections precisely because $\WD{ij} =0$ here.

\section  {Type D vacuum spacetimes in five dimensions}
\label{D5D}

Let us now study the five-dimensional case.
{Note that the algebraic relation (\ref{Wcomptsid}) between $-2\WDS{ij} $ and $C_{ijkl}$ is equivalent to the relation between the Ricci and the Riemann tensor of a $m-2$ dimensional space. Therefore} in five dimensions $C_{ijkl}$
is equivalent to $\WDS{ij} $ and thus a type D  Weyl tensor in five dimesions is fully determined by $ \WD{ij} $. In fact, for $n=5$ it is possible to solve the second constraint from (\ref{Wcomptsid}) for $C_{ijkl}$:
\be
\fl C_{ijkl} \eq5d 2 \left(\kd{il} \WDS{jk} -  \kd{ik} \WDS{jl} -
\kd{jl} \WDS{ik} + \kd{jk} \WDS{il}  \right) - \WD{} \left(
\kd{il} \kd{jk} - \kd{ik} \kd{jl}   \right). \label{Weyl5D} \ee
Thus in the five dimensional case the algebraic equations  we consider, (\ref{Bia8}), (\ref{algcomb15}),
(\ref{eqB15i=k}), (\ref{eqB12B5sym}),
 can be expressed in terms of $\WD{ij} $,
$L_i$, and $L_{ij}$.
Plugging (\ref{Weyl5D}) into (\ref{Bia8}), recalling equation   (\ref{Bia8a4}) and contracting with $L_k$ one  finds the equation
\be
L\WDS{ij} +2\WD{} L_iL_j-\WD{} L\delta_{ij}=0. \label{5dgeod}
\ee
For $n=5$ equation   (\ref{algcomb15})  takes the form
\bea
\fl 0
= \WDA{jk} L_{im} + ( \WDA{im} + 3\WDS{im} ) A_{kj} + \WDA{mj}
L_{ik} + (\WDA{ik} +3 \WDS{ik} ) A_{jm} + \WDA{km} L_{ij}
\nonumber\\ \fl + (\WDA{ij} +3\WDS{ij} ) A_{mk} + \delta_{ij} (
\WDS{ms} L_{sk} - \WDS{ks} L_{sm} ) + \delta_{ik} ( \WDS{js}
L_{sm} - \WDS{ms} L_{sj} ) \nonumber\\ \fl + \delta_{im} (
\WDS{ks} L_{sj} - \WDS{js} L_{sk} ) + \WD{ } [\delta_{ij}
A_{km}+\delta_{ik} A_{mj}+\delta_{im} A_{jk}] . \label{B15}
\eea
Equation (\ref{eqB15i=k}) reduces to
\bea
0=\WDA{mj} S+2\WD{ } A_{jm}
+\WDA{ji} (S_{im}+2A_{im})+\WDA{im}  (S_{ij}+2A_{ij}) \nonumber \\
+\WDS{ji} (S_{im}-2A_{im}) +\WDS{mi} (-S_{ij}+2A_{ij}) ,
\label{eqB15i=k5d} \eea and
equation   (\ref{eqB12B5sym}) has the form
\be
 \!\!\!\!\!\!\!\!\!\!\!\!
3[(\WDS{ij} +\WDA{ij} ) S_{jk} + (\WDS{kj} +\WDA{kj} ) S_{ji} -S \WDS{ki} ]
=\delta_{ik} (2\WDS{js} S_{js} -\WD{ } S).\label{b5b12sym5d}
\ee
In the following sections we  study (non-)geodecity of multiple WANDs
(section   \ref{geod5D}),  spacetimes admitting non-twisting WANDs $A_{ij}=0$ (section   \ref{5Dnontwist}) and
spacetimes with $\WDA{ij} =0$
(section   \ref{secPhiA=0}).

\subsection  {Geodeticity of multiple WANDs}
\label{geod5D}

It is interesting to return now to equation (\ref{Bia8}), which is related to the (non-)geodetic character of multiple WANDs and in five dimensions implies
(\ref{5dgeod}). Since we already know from Proposition~\ref{propgeodetic} that WANDs are necessarily geodetic when $\WDA{ij} \neq 0$, let us focus here on the case $\WDA{ij} =0$.
If $\WD{} =0$ we see that either $L=0$ or $\WDS{ij} =0$, the latter case being now a conformally flat spacetime. Therefore an $n=5$ type D Einstein spacetime requires ($\WDA{ij} =0$ and) $\WD{} \neq 0$ in order to admit a non-geodetic multiple WAND. In this case it follows from (\ref{5dgeod}) that there exists an eigenframe of $\WDS{ij} $ such that
\be
\WDS{ij} =\WD{ } \mbox{diag} ( 1,1,-1), \qquad L_2=L_3=0 , \label{nongeod5}
\ee
so that $L_4\neq 0$ is responsible for the WAND $\bl$ being non-geodetic.
Such spacetime is necessarily shearing since
the ``canonical'' form of $\WDS{ij} $ given in equation   (\ref{nongeod5})  is not compatible with that of
equation   (\ref{Phidelta}).
It would be interesting to find such five dimensional vacuum type D spacetime with a non-geodetic WAND
or prove that such spacetime does not exist.

{To summarize,}

\begin{proposition}
\label{propnongeod5D}
In five dimensions,
the only type D spacetimes with non-geodetic multiple WAND $\bl$ are those
satisfying
$\WDA{ik} =0 $ and $\WDS{ik} \not=0$, $\WDS{ik} =\mbox{diag}\{\WD{} ,\ \WD{} ,\ -\WD{} \}$.
\end{proposition}

\subsection  {``Non-twisting'' case - $A_{ij}=0$}
\label{5Dnontwist}

In the non-twisting case $A_{ij}=0$, equation (\ref{eqB15i=k5d})
reduce to
\be
 \WD{ji} S_{im} - \WD{mi} S_{ij} +\WDA{mj} S =0 \label{b15Anul} .
\ee

Now we can, without loss of generality,  choose a frame in which the symmetric matrix $S_{ij}$
is diagonal
\be
S_{ij}=\mbox{diag} ( s_{(2)},s_{(3)},s_{(4)} ).
\ee
Then equations (\ref{b15Anul}) and  (\ref{b5b12sym5d}) take the form (recall that we do not sum over indices in brackets)
\bea
\WDS{ik} (s_{(k)} - s_{(i)}) +\WDA{ik} (s_{(k)}+s_{(i)}-S)&=&0,\nonumber\\
\WDS{ik} (s_{(k)}+s_{(i)}-S) +\WDA{ik} (s_{(k)}-s_{(i)})&=& {\textstyle\frac{1}{3}}\delta_{ik} (2\WDS{js} S_{js} -\WD{ } S) . \label{rceAnul}
\eea

Now let us study components of the two above equations for $i\not=k$. By summing the two above equations we get
\be
(2s_{(k)} -S) (\WDS{ik} +\WDA{ik} )=0 \qquad (i\not=k) .
\ee
In the ``generic'' case with $2s_{(i)} \not=S \ \forall i$, this implies
\be
\WDA{ik} =0= \WDS{ik} \ \ \mbox{for}\ \ i\not= k .
\label{resultAnulPA}
\ee
Consequently, $\WDS{ij} $ is also diagonal and from equation   (\ref{rceAnul})
\be
\WDS{ij} =\mbox{diag} ( p_{(2)},p_{(3)},p_{(4)} ), \ \ \ p_{(i)}=\frac{2\WDS{js} S_{js} -\WD{ } S}{3(2s_{(i)}-S)}.
\label{resultAnulPS}
\ee
Using (\ref{resultAnulPS}),
it is straightforward to express (two of) the $p_{(i)}$ in terms of the $s_{(i)}$ solving the linear relations (which are not all independent):
\bea
(s_{(2)}-s_{(3)}-s_{(4)})p_{(2)}&=&(-s_{(2)}+s_{(3)}-s_{(4)})p_{(3)},\\
(s_{(2)}-s_{(3)}-s_{(4)})p_{(2)}&=&(-s_{(2)}-s_{(3)}+s_{(4)})p_{(4)},\\
(-s_{(2)}+s_{(3)}-s_{(4)})p_{(3)}&=&(-s_{(2)}-s_{(3)}+s_{(4)})p_{(4)}.
\eea
Thus
\begin{proposition}
\label{prop5Da}
In five dimensions,
in the ``generic'' {($2s_{(i)}  \not=S \ \forall i$)} non-twisting ($A_{ij}=0$) type D spacetime,  $\WDA{ij} $ also vanishes and
$\WDS{ij} $ can be diagonalized together with $S_{ij}$.
\end{proposition}
Note that special cases   with  $2s_{(i)}=S$ for some $i$
have to be treated separately:\\
1) If one of $s_{(i)}=S/2$, e.g. $s_{(4)}=S/2$, and the others differ from $S/2,0$ then
only $\WDS{44} \not= 0$, all other component of $\WDS{ij} =0$ and $\WDA{ij} =0$.
\\
2) If e.g. $s_{(2)}=s_{(3)}=S/2$, $s_{(4)}=0$ then
$\WDS{24} =\WDS{34} =\WDS{44} =\WDA{24} =\WDA{34} =0 $, the other components ($\WDS{22} $,
$\WDS{33} $, $\WDS{23} $, $\WDA{23} $) are arbitrary.

\subsection  {Case $\WDA{ij} =0$}
\label{secPhiA=0}

For $\WDA{ij} =0$
 equations (\ref{eqB15i=k5d}), (\ref{DPhi2_as})   and (\ref{b5b12sym5d}) take the form
\bea
&&2(\WDS{mi} A_{ij} - \WDS{ji} A_{im} +\WD {} A_{jm})
+\WDS{ji} S_{im} - \WDS{mi} S_{ij} =0 ,\label{b15PAnul}\\
&&-\WDS{im} A_{ij} + \WDS{ji} A_{im} +2\WD {} A_{jm}
+\WDS{ji} S_{im} - \WDS{mi} S_{ij} =0,\label{b3PAnul}\\
&&3(\WDS{ij} S_{jk} + \WDS{kj} S_{ji} -S \WDS{ki} )=\delta_{ik} (2\WDS{jl} S_{jl} -\WD{ } S).\label{b5b12PAnul}
\eea

In previous section   \ref{5Dnontwist} it was efficient to choose a frame in which $S_{ij}$ was diagonal,
however, now it is more efficient to choose a frame in which  $\WDS{ij} $ is diagonal,
$\WDS{ij} =\mbox{diag} \{ p_{(2)},\ p_{(3)},\ p_{(4)}\}$.
  Then
we obtain from (\ref{b15PAnul})--(\ref{b5b12PAnul}) the following set of equations
\bea
&&(2p_{(m)}+2p_{(j)}-2\WD{} ) A_{mj} +S_{mj} (p_{(j)}-p_{(m)})=0,\label{b15PAnulPSdig}\\
&&(-p_{(m)}-p_{(j)}-2\WD{} ) A_{mj} +S_{mj} (p_{(j)}-p_{(m)})=0,\label{b3PAnulPSdiag}\\
&&3(p_{(i)}+p_{(k)})S_{ik}=\delta_{ik} (3Sp_{(i)}+2\WDS{jl} S_{jl} -\WD{} S).\label{b5b12PAnulPSdiag}
\eea
In the ``generic'' case $p_{(i)}+p_{(k)} \not=0$, $\forall i,\ k$, from equation   (\ref{b5b12PAnulPSdiag})
\be
S_{ik}=\mbox{diag}\{ s_{(2)},\ s_{(3)},\ s_{(4)}\}, \ \ \ s_{(i)}=\frac{S}{2}
+\frac{2\WDS{jl} S_{jl} -\WD{} S}{6p_{(i)}}.\label{PA0-S}
\ee
From (\ref{PA0-S}) we get the relations (which can be solved to fix two of the $s_i$, if desired):
\bea
s_{(2)}p_{(3)}(p_{(2)}+p_{(4)})&=&s_{(3)}p_{(2)}(p_{(3)}+p_{(4)}),\\
s_{(2)}p_{(4)}(p_{(2)}+p_{(3)})&=&s_{(4)}p_{(2)}(p_{(3)}+p_{(4)}),\\
s_{(3)}p_{(4)}(p_{(2)}+p_{(3)})&=&s_{(4)}p_{(3)}(p_{(2)}+p_{(4)}).
\eea
Subtracting (\ref{b15PAnulPSdig}) and (\ref{b3PAnulPSdiag}) we obtain $(p_{(m)}+p_{(j)})A_{mj}=0$
and thus in the ``generic'' case $p_{(m)}+p_{(j)} \not=0$, $\forall m,j$,
\be
A_{mj}=0.
\ee

\begin{proposition}
\label{prop5Db}
In five dimensions, the multiple WAND $\bl$
in a ``generic'' ($p_{(i)}+p_{(j)} \not=0$, $\forall i,j$)
type D spacetime with $\WDA{ik} =0 $ and $\WDS{ik} \not=0$, is geodetic and non-twisting
 ($A_{ij}=0$) and  $\WDS{ik} $  and $S_{ij}$ can be diagonalized  together.
\end{proposition}

There are some special cases to be treated:\\
- Case a) one $p_{(i)}=0$ and $\WD{} \not=0$: without loss of generality we choose $p_{(2)}=0$, then
from (\ref{b15PAnulPSdig})--(\ref{b5b12PAnulPSdiag}) $2\WDS{jl} S_{jl} -\WD{} S=0$,
${ S}_{ij}=\mbox{diag}\{ 0,S/2,S/2\}$, $A_{mj}=0$.\\
- Case b) only one $p_{(i)}\not=0$:  without loss of generality we choose $p_{(4)}\not=0$, $p_{(2)}=p_{(3)}=0$ then from (\ref{b15PAnulPSdig})--(\ref{b5b12PAnulPSdiag}) $2\WDS{jl} S_{jl} -\WD{} S=0$,
 $s_{(2)}+s_{(3)}=s_{(4)}=S/2$ and $S_{23}$ is arbirary,
$A_{ij}$ vanishes.\\
- Case c) only one pair satisfies $p_{(m)}+p_{(j)} =0$, $p_{(j)}\not=0$:  without loss of generality we choose
$p_{(3)}+p_{(4)}=0$, i.e. $p_{(2)}=\WD{} $, then
the diagonal components of $S_{ij}$ still satisfy (\ref{PA0-S}),
from (\ref{b15PAnulPSdig})--(\ref{b5b12PAnulPSdiag}) $S_{34}$ is arbitrary and
\be
(p_{(m)}+p_{(j)})A_{mj}=0, \ \ \ 2\WD{} A_{mj}=(p_{(j)}-p_{(m)})S_{mj}\label{PA0eqA}
\ee
and thus if $\WD{} \not=0$, $A_{34}=-\frac{p_{(3)}}{\WD{} }S_{34}$.
If $\WD{} =0$, then $S_{34}=0$ and ${ S}_{ij}$ is diagonal and $A_{23}$ is arbitrary.
\\
- Case d) two pairs satisfy $p_{(m)}+p_{(j)} =0$: without loss of generality we choose
$p_{(2)}=p_{(3)}=-p_{(4)}=\WD{} $. From (\ref{PA0-S}) it follows that
the diagonal components of $S_{ij}$, $s_{(2)}$ and $s_{(3)}$, vanish  and  $s_{(4)}$ is arbitrary.
Equation   (\ref{b5b12PAnulPSdiag}) implies that $S_{24}$ and $\ S_{34}$ are arbitrary and from
equation   (\ref{PA0eqA}) we get
$A_{23}=0,\ \ A_{24}=-S_{24}, \ \ \ A_{34}=-S_{34}$.
This case is the non-geodetic case (\ref{nongeod5})
from section   \ref{geod5D}.

\subsection  {An example - Myers-Perry black hole}

\label{subsec_MP}

As an illustrative example we give $S_{ij} $, $A_{ij} $, $\WDS{ij}
$ and $\WDA{ij} $ for the five-dimensional Myers-Perry black hole
\cite{MyePer86} \beah \fl \ \ \ \ \ {\rm d}s^2=\frac{\rho^2}{4
\Delta} {\rm d}x^2 + \rho^2 {\rm d} \theta^2 -{\rm d}t^2 + (x+a^2)
\sin^2 \theta {\rm d} \phi^2 + (x+b^2) \cos^2 \theta {\rm d}
\psi^2  \\ + \frac{{r_0}^2}{\rho^2} ({\rm d}t + a \sin^2 \theta
{\rm d} \phi + b \cos^2 \theta {\rm d} \psi)^2 , \eeah where $$
\rho^2=x+a^2 \cos^2 \theta + b^2 \sin^2 \theta ,\ \
\Delta=(x+a^2)(x+b^2)-{r_0}^2 x . $$ Two (multiple, geodetic)
WANDs (related by reflection symmetry) are given by {\cite{FroSto03}}
\be
\fl
\bl = \frac{(x+a^2)(x+b^2)}{\Delta} \left[ \partial_t - \frac{a}{x+a^2} \partial_{\phi} - \frac{b}{x+b^2} \partial_{\psi}  \right] +
 2 \sqrt{x} \partial_x,
\ee
\be
\fl
\bn = \alpha \left( \frac{(x+a^2)(x+b^2)}{\Delta} \left[ \partial_t - \frac{a}{x+a^2} \partial_{\phi} - \frac{b}{x+b^2} \partial_{\psi}  \right] -
 2 \sqrt{x} \partial_x \right),
\ee
where we chose $\alpha = -\Delta / {2 \rho^2 x}$ in order to satisfy the normalization condition $\bl \cdot \bn = 1$.

As a basis of spacelike vectors we choose three eigenvectors of $S_{ij}$
\bea
\bm{2} &=& \frac{1}{\rho} \partial_{\theta}, \nonumber \\
\bm{3} &=& \frac{1}{\sqrt{x} \chi} \left(- a b \partial_t + b \partial_{\phi} + a \partial_{\psi} \right), \\
\bm{4} &=& \frac{1}{\rho \chi} \left[ (a^2-b^2) \sin \theta \cos \theta \partial_t
      - a \tan^{-1} \theta  \partial_{\phi} + b \tan \theta \partial_{\psi} \right], \nonumber
\eea
with $\chi=\sqrt{a^2 \cos^2 \theta + b^2 \sin^2 \theta}$.
In this frame
\be
S_{ij}= \left( \begin {array}{ccc} {\frac {\sqrt {x}}{{\rho}^{2}}}&0&0
\\\noalign{\medskip}0& \frac {1}{\sqrt {x}}&0
\\\noalign{\medskip}0&0&{\frac {\sqrt {x}}{{\rho}^{2}}}\end {array}
 \right),
\ \ \ \
A_{ij}=  \frac{\chi}{\rho^2}  \left( \begin {array}{ccc}
0&0&-1\\
0&0&0\\
1&0&0\end {array}
 \right)  ,
\ee
and
\be
 \WDS{ij} = \frac{{r_0}^2}{\rho^4} \left( \begin {array}{ccc} {\frac {\rho^2- 2 x}{{\rho}^{2}}}&0&0
\\\noalign{\medskip}0& -1 &0
\\\noalign{\medskip}0&0&{\frac {\rho^2- 2 x}{{\rho}^{2}}}\end {array}
 \right),
\ \ \ \
\WDA{ij} =  \frac{2 {r_0}^2 \chi \sqrt{x} }{\rho^6}  \left( \begin {array}{ccc}
0&0&1\\
0&0&0\\
-1&0&0\end {array}
 \right) .
\ee

Notice that in the static (Schwarzschild) limit ($a=0=b$ so that $\rho^2=x$) one has $S_{ij}=\delta_{ij}/\sqrt{x}$ and $\sigma_{ij}=0=A_{ij}$, and indeed for $\WD{ij} $ we recover the form discussed in subsection~\ref{shearfreeD} in the shearfree expanding case and in subsection~\ref{5Dnontwist} in the ``generic'' non-twisting case (with $p_{(2)}=p_{(3)}=p_{(4)}$).

\section  {Discussion}
\label{discussion}

Let us finally outline main results presented in the paper.

In the first part of the paper  (Sections \ref{GID} and \ref{warped})
we study constraints on Weyl types of a spacetime following from
various  assumptions on  geometry.
It turns out that:\\
- Static spacetimes are of types G, ${\rm I}_i$, D or conformally flat (Proposition \ref{propstatic}). \\
- ``Expanding'' stationary spacetimes with appropriate reflection symmetry
    belong to these types as well (Proposition \ref{propstationary}). \\
- Warped spacetimes with   one-dimensional Lorentzian
factor are again of types G, ${\rm I}_i$, D and O  (Proposition \ref{prop_warped1}). \\
- Warped spacetimes with two-dimensional Lorentzian factor are
necessarily of types D or O (Proposition \ref{prop_warped2}), in particular this also applies  to spherically symmetric
spacetimes (Proposition \ref{prop_spherical}).

 These results may  have useful practical applications in determining the algebraic type of specific spacetimes (or at least in ruling out some types) just by ``inspecting'' the given metric and without performing any calculations. This is particularly important in higher dimensions, where it is more difficult to determine the algebraic class of a given metric.

In the second part of the
paper (sections \ref{DHD} and \ref{D5D}) we study properties of type D vacuum spacetimes
in general (without assuming that the spacetime is static, stationary or warped).
In five dimensions a type D Weyl tensor is determined by a $3 \times 3$
 matrix $\WD{ij} $ with symmetric and antisymmetric parts being $\WDA{ij} $ and $\WDS{ij} $,
 respectively.
In general in the non-twisting case $\WD_{ij} $ is symmetric while in the twisting case antisymmetric
  part  $\WDA{ij} $ appears.
In higher dimensions $n>5$ the $(n-2) \times (n-2)$ matrix  $\WD{ij} $ does not
 contain complete information about the Weyl tensor, but it still plays an important role.
 The matrix $\WD{ij} $ can  also be used for further
 classification of type D or II spacetimes, for example according to
possible degeneracy of eigendirections of $\WD{ij} $. Special classes are
also cases with $\WD{ij} $ being symmetric or  vanishing (such examples for $n \geq 7$ are given in section   \ref{nongeod}) etc.

First we focused on the geodeticity of multiple WANDs in type D vacuum spacetimes (these
are always geodetic for $n=4$).
It was shown that:\\
- The multiple WAND in a vacuum spacetime is geodetic in the ``generic'' case, i.e.  if
 $\WDA{ij} \not=0 $
or if all eigenvalues of $\WDS{ij} $
are distinct from minus the trace of $\WD{ij} $  (Proposition \ref{propgeodetic}).\\
- It is also geodetic in the type D, shearfree case whenever $\WD{ij} \not=0$  (Proposition \ref{propshearfree}).\\
- However, explicit examples of vacuum type D spacetimes with non-geodetic multiple WAND in $n\geq 7$ dimensions are given in section   \ref{nongeod}.
   This provides us with the first examples of spacetimes ``violating'' the geodetic part of the Goldberg-Sachs theorem. \\
- In five dimensions multiple WANDs are also geodetic when $\WDA{ij} =0$ and $\WDS{ij} \not=0 $ has a ``generic'' form (Proposition \ref{prop5Db}), special cases are discussed in section   \ref{secPhiA=0}.\\

Properties of the matrix $\WD{ij} $, as well as the expansion and twist matrices $S_{ij}$ and $A_{ij}$
have been also studied:\\
- For warped spacetimes with a one/two-dimensional Lorentzian factor
(thus also for static spacetimes)
the antisymmetric part of $\WD{ij} $, $\WDA{ij} $, vanishes.\\
- In  vacuum type D spacetimes  admitting a shearfree expanding WAND,
$\WDS{ij} $ is proportional to $\delta_{ij}$ and
if $A_{ij}=0$ (this always holds in odd dimensions \cite{OrtPraPra07})  then  $\WDA{ij} =0$
and in the case with $\WDS{ij} \not=0$ also vice versa
(Proposition \ref{propshearfree2}).\\
- In five dimensions in a ``generic'' Einstein type D non-twisting
spacetime,  $\WDA{ij} $ vanishes and eigendirections of $\WD{ij} $  coincide with those of $S_{ij}$ (Proposition \ref{prop5Da}).\\
- In five dimensions in a ``generic'' vacuum type D
spacetime with symmetric  $\WD{ij} $, the multiple WAND $\bl$ is  non-twisting and eigendirections of $\WD{ij} $  and $S_{ij}$ coincide (Proposition \ref{prop5Db}).\\

These results provide interesting connections between geometric properties of principal null
congruences and Weyl curvature. Hopefully, they can be also used for constructing exact
type D solutions with particular properties.

\ack
V.P. and A.P. acknowledge support from research plan No
AV0Z10190503 and research grant KJB100190702.

\appendix
\section{Optics of WANDs in Kerr-NUT-AdS spacetimes in arbitrary dimension}

As discussed in Sec. \ref{sub_sec_limitations}, the assumption about non-zero ``expansion'' in Proposition \ref{propstationary} is essential. In this appendix we study optical properties of WANDs in Kerr-NUT-AdS spacetimes in arbitrary dimension \cite{CheLuPop06} and show that the ``expansion'' in these cases is always non-vanishing.
These metrics are thus subject to Proposition 2. Indeed,
it has been already shown {in \cite{Hamamotoetal06}} that these spacetimes are of
type D. In addition, since the expansion is non-zero, we can expect that possible
(still stationary) generalizations of these spacetimes (such as charged black holes)
with appropriate reflection symmetry are of types G, $I_i$ or D (see also footnotes
on page 5).
This appendix also extends our example of five-dimensional Myers-Perry given in Section  \ref{subsec_MP} to the case with NUT parameters and cosmological constant and to arbitrary dimension.  Note, however, that
now we use convenient but  physically less ``transparent'' coordinates $(x_1,\dots ,x_m, \psi_0, \dots \psi_{m-1} )$ in even dimensions $n=2 m $ and
 $(x_1,\dots ,x_m, \psi_0, \dots \psi_{m} )$ in odd dimensions $n=2 m + 1$, introduced in  \cite{CheLuPop06}.
In our calculations, we employ results obtained in \cite{Hamamotoetal06}.

The metric { of \cite{CheLuPop06}} for { even and odd} dimensions is{, respectively,} \\
$n=2 m $:
\be
\rmd s^2 = \sum_{\mu=1}^{m} \frac{\rmd x_{\mu}^2}{Q_\mu}+
\sum_{\mu=1}^{m} Q_{\mu} \left( \sum_{k=0}^{m-1} A_{\mu}^{(k)}
\rmd \psi_k \right)^2, \label{KNUTeven}
\ee
$n=2 m + 1$:
\begin{equation}
\rmd s^2 = \sum_{\mu=1}^{m} \frac{\rmd x_{\mu}^2}{Q_\mu}+
\sum_{\mu=1}^{m} Q_{\mu} \left( \sum_{k=0}^{m-1} A_{\mu}^{(k)}
\rmd \psi_k \right)^2+{\tilde S}\left( \sum_{k=0}^{m} A^{(k)}
\rmd \psi_k \right)^2. \label{KNUTodd}
\end{equation}
{ The functions $Q_{\mu}$, $A_{\mu}^{(k)}$, $A^{(k)}$ and $\tilde S$ depend only on the coordinates $(x_1,\dots ,x_m)$ and their explicit expressions} are given in \cite{CheLuPop06,Hamamotoetal06}.

\subsection{Even dimensions, $n=2 m $}

An orthonormal frame of 1-forms
$\{ {\ble}^{(A)} \}=\{ \ble^{(\mu)}, \ble^{(m+\mu)} \}$ with $A=1,2,\dots 2m$, $\mu=1,2,\dots , m$,
\begin{equation}
\ble^{(\mu)} = \frac{\rmd x_{\mu}}{\sqrt{Q_{\mu}}}, \quad
\ble^{(m+\mu)} = \sqrt{Q_{\mu}}
\left( \sum_{k=0}^{m-1} A_{\mu}^{(k)} \rmd \psi_k \right)\label{frame_forms_even}
\end{equation}
was intoduced in \cite{Hamamotoetal06}.
{ Denoting the duals of these forms with lower indices}, let us here also define a null frame { of vectors} $\bl$, $\bn$, $\bmd{i} $ by
\be
\bl=\frac{{\rm i}}{\sqrt{2Q_m}}(\ble_{(m)}+{\rm{i}}\ble_{(2m)}),\ \ \
\bn=-{\rm i}\sqrt{\frac{Q_m}{2}}(\ble_{(m)}-{\rm{i}}\ble_{(2m)}), \label{nullvectors}
 %\bmd{i} =\ble_{(\mu)},\ \ble_{(m+\mu)}\ \ \ \mu=1\dots m-1,\ \ i=2\dots n
\ee
with $\bmd{i} $ ($i=2\dots n-1$) corresponding to $\ble_{(\mu)}$, $\ble_{(m+\mu)}$ ($ \mu=1\dots m-1$  from now on).
{ One can show \cite{Hamamotoetal06} that the null vectors $\bl$, $\bn$  are
 multiple WANDs of the type~D metric (\ref{KNUTeven}) and that they are geodetic (and affinely parametrized)}.
{ Both WANDs are complex in the coordinates used above, but note that they become in fact real in ``physical'' coordinates} since
the metric (\ref{KNUTeven}) was obtained from a real Lorentzian metric by a Wick rotation with
 $x_m={\rm i}r$ in \cite{CheLuPop06} and $Q_m<0$ { in the outer stationary region, where}  $\partial / \partial r$ { is spacelike}.  Thus $\sqrt{Q_m}={\rm i}\sqrt{|Q_m|}$, { so that reintroducing $r$,  both vectors ${\rm i}\ble_{(2m)}$
and $\ble_{(m)}$  become real  ($e_{(m)}^\alpha (r)={\rm i}\delta^{\alpha m}\sqrt{Q_m}$)}.

Let us now express the matrix $L_{ij}$ {(defined in section~\ref{prelim})} in terms of Ricci rotation coefficients, which can be easily obtained from the connection 1-forms given in \cite{Hamamotoetal06}
\be
\fl
L_{ij}=\ell_{a;b}m^a_{(i)}m^b_{(j)}=\frac{1}{\sqrt{2Q_m}}(e_{(m)a;b}+{\rm{i}} e_{(2m)a;b}) m^a_{(i)}m^b_{(j)}
%=\frac{1}{\sqrt{2Q_m}}(\Mm_{ij}+{\rm{i}}\Mmm_{ij})
=-\frac{1}{\sqrt{2Q_m}}(\gamma^m_{\ \ ij}+{\rm{i}}\gamma^{2m}_{\ \ ij}), \label{Lijeven}
\ee
with
\bea
\gamma^m_{\ \ \mu\mu}&=&\gamma^m_{\ \ m+\mu\ m+\mu}=-\frac{x_m\sqrt{Q_m}}{x_m^2-x_\mu^2},\\
%\gamma^m_{\ 2m+1\ 2m+1}&=&-\frac{\sqrt{Q_m}}{x_m},\\
\gamma^{2m}_{\ \ m+\mu\ \mu}&=&-\gamma^{2m}_{\ \ \mu\ m+\mu}=-\frac{x_\mu\sqrt{Q_m}}{x_m^2-x_\mu^2}, \label{Riccieven}
\eea
 and with  remaining Ricci rotation coefficients entering (\ref{Lijeven}) being zero. %Note also that $\Mi_{jk} = -\gamma^i_{\ jk} $.
Then
\be
\fl
S_{ij}=\frac{r}{\sqrt{2}} \left( \begin {array}{cc} {\delta_{\mu\nu}}\frac{1}{r^2+x_\mu^2}&0
\\\noalign{\medskip}0&\delta_{\mu\nu} \frac{1}{r^2+x_\mu^2}
%\\\noalign{\medskip}0&0& 1
\end {array}
 \right),
\ \ \ \
A_{ij}=  \frac{1}{\sqrt{2}} \left( \begin {array}{cc}
0&-\delta_{\mu\nu}\frac{x_\mu}{r^2+x_\mu^2}\\
\delta_{\mu\nu}\frac{x_\mu}{r^2+x_\mu^2}&0\end {array}
 \right),
\ee
{ where terms proportional to $\delta_{\mu\nu}$ symbolically represent a $(m-1)\times (m-1)$ diagonal block.}
{ Note that $S_{ij}\propto\delta_{ij}$ (that is, the shear is zero) iff $n=4$}. From this form of $S_{ij}$ it follows that shear is non-zero for arbitrary even dimension  $n>4$ and
expansion
\be
S=\sqrt{2}r\sum_{\mu=1}^{m-1} \frac{1}{r^2+x_\mu^2}
\ee
is non-zero in arbitrary even dimension $n \geq 4$. { Note indeed that the  WANDs $\bl$ and $\bn$ are related by reflection symmetry, in agreement with the discussion in section~\ref{GID}.}
{The twist is also obviously non-zero for any $n \geq 4$. Recall \cite{CheLuPop06} finally that for $n=4$ the metric~(\ref{KNUTeven}) represents a subclass of the Pleba\'nski-Demia\'nski family of type D spacetimes with two expanding, twisting and non-shearing principal null directions \cite{Stephanibook}.}

\subsection{Odd dimensions, $n=2 m + 1$}

In odd dimensions, in addition to (\ref{frame_forms_even}) we define
\begin{equation}
\ble^{(2m+1)} = \sqrt{\tilde S}
\left( \sum_{k=0}^{m} A^{(k)} \rmd \psi_k \right).
\end{equation}
Then the null frame consists of $\bl$, $\bn$ given in (\ref{nullvectors}),
 $\bmd{i} $ ($i=2\dots n-1$) corresponding to $\ble_{(\mu)}$, $\ble_{(m+\mu)}$ ($ \mu=1\dots m-1$), and $\ble_{(2m+1)}$. { Again, the null vectors $\bl$ and $\bn$  are geodetic multiple WANDs of the type D metric (\ref{KNUTodd}) \cite{Hamamotoetal06}}.

Now together with (\ref{Riccieven}) we have
\be
%\gamma^m_{\ \ \mu\mu}&=&\gamma^m_{\ \ m+\mu\ m+\mu}=-\frac{x_m\sqrt{Q_m}}{x_m^2-x_\mu^2},\\
\gamma^m_{\ 2m+1\ 2m+1}=-\frac{\sqrt{Q_m}}{x_m},\\
%\gamma^{2m}_{\ \ m+\mu\ \mu}&=&-\gamma^{2m}_{\ \ \mu\ m+\mu}=-\frac{x_\mu\sqrt{Q_m}}{x_m^2-x_\mu^2},
\ee
and thus
\be
\fl
S_{ij}=\frac{1}{\sqrt{2}r} \left( \begin {array}{ccc} {\delta_{\mu\nu}}\frac{r^2}{r^2+x_\mu^2}&0&0
\\\noalign{\medskip}0&\delta_{\mu\nu} \frac{r^2}{r^2+x_\mu^2}&0
\\\noalign{\medskip}0&0& 1
\end {array}
 \right),
\ \ \ \
A_{ij}=  \frac{1}{\sqrt{2}} \left( \begin {array}{ccc}
0&-\delta_{\mu\nu}\frac{x_\mu}{r^2+x_\mu^2}&0\\
\delta_{\mu\nu}\frac{x_\mu}{r^2+x_\mu^2}&0&0\\
0&0&0\end {array}
 \right)  .
\ee
Shear, expansion { and twist} are thus non-zero for arbitrary odd dimension  $n>4$.

\section  *{References}


\begin{thebibliography}{10}

\bibitem{Stephanibook}
H.~Stephani, D.~Kramer, M.~MacCallum, C.~Hoenselaers, and E.~Herlt.
\newblock {\em Exact Solutions of {E}instein's Field Equations}.
\newblock Cambridge University Press, Cambridge, second edition, 2003.

\bibitem{Coleyetal04}
A.~Coley, R.~Milson, V.~Pravda, and A.~Pravdov\'a.
\newblock Classification of the {W}eyl tensor in higher dimensions.
\newblock {\em Class. Quantum Grav.}, 21:L35--L41, 2004.

\bibitem{Milsonetal05}
R.~Milson, A.~Coley, V.~Pravda, and A.~Pravdov\'a.
\newblock Alignment and algebraically special tensors in {L}orentzian geometry.
\newblock {\em Int. J. Geom. Meth. Mod. Phys.}, 2:41--61, 2005.

\bibitem{ColPel06}
A.~Coley and N.~Pelavas.
\newblock Algebraic classification of higher dimensional spacetimes.
\newblock {\em Gen. Rel. Grav.}, 38:445--461, 2006.

\bibitem{OrtPra06}
M.~Ortaggio and V.~Pravda.
\newblock Black rings with a small electric charge: gyromagnetic ratios and
  algebraic alignment.
\newblock {\em JHEP}, 12:054, 2006 [gr-qc/0609049]

\bibitem{IdaUch03}
D.~Ida and Y.~Uchida.
\newblock Stationary {E}instein-{M}axwell fields in arbitrary dimensions.
\newblock {\em Phys. Rev. {\rm D}}, 68:104014, 2003.

\bibitem{FroSto03}
V.~P. Frolov and D.~Stojkovi\'c.
\newblock Particle and light motion in a space-time of a five-dimensional
  rotating black hole.
\newblock {\em Phys. Rev. {\rm D}}, 68:064011, 2003.

\bibitem{Pravdaetal04}
V.~Pravda, A.~Pravdov\'a, A.~Coley, and R.~Milson.
\newblock Bianchi identities in higher dimensions.
\newblock {\em Class. Quantum Grav.}, 21:2873--2897, 2004.

\bibitem{MyePer86}
R.~C. Myers and M.~J. Perry.
\newblock Black holes in higher dimensional space-times.
\newblock {\em Ann. Phys. (N.Y.)}, 172:304--347, 1986.

\bibitem{OrtPraPra07}
M.~Ortaggio, V.~Pravda, and A.~Pravdov\'a.
\newblock Ricci identities in higher dimensions.
\newblock {\em Class. Quantum Grav.}, 24:1657--1664, 2007.

\bibitem{PraPra05}
V.~Pravda and A.~Pravdov\'a.
\newblock {WAND}s of the black ring.
\newblock {\em Gen. Rel. Grav.}, 37:1277--1287, 2005.

\bibitem{petrov}
A.~Z. Petrov.
\newblock {\em Einstein Spaces}.
\newblock Pergamon Press, Oxford, translation of the 1961 {R}ussian edition,
  1969.

\bibitem{Kinnersley:JMP69}
W. Kinnersley.
\newblock Type {D} vacuum metrics.
\newblock {\em J. Math. Phys.}, 10:1195--1203, 1969.

\bibitem{PraPraOrt07KS}
V.~Pravda, A.~Pravdov\'a, and M.~Ortaggio,
\newblock in preparation.

\bibitem{Hamamotoetal06}
N.~Hamamoto, T.~Houri, T.~Oota, and Y.~Yasui.
\newblock Kerr-{NUT}-de~{S}itter curvature in all dimensions.
\newblock {\em J. Phys. A}, 40:F177--F184, 2007.

\bibitem{CheLuPop06}
W.~Chen, H.~L{\"u}, and C.~N. Pope.
\newblock General {K}err-{NUT}-{AdS} metrics in all dimensions.
\newblock {\em Class. Quantum Grav.}, 23:5323--5340, 2006.

\bibitem{EmpRea02prl}
R.~Emparan and H.~S. Reall.
\newblock A rotating black ring solution in five dimensions.
\newblock {\em Phys. Rev. Lett.}, 88:101101, 2002.

\bibitem{Harmark04}
T.~Harmark.
\newblock Stationary and axisymmetric solutions of higher-dimensional general
  relativity.
\newblock {\em Phys. Rev. {\rm D}}, 70:124002, 2004.

\bibitem{ElvangFiguerasSaturn}
H. Elvang and P. Figueras.
\newblock Black saturn, hep-th/0701035.

\bibitem{Pomeransky2006}
A.~A. Pomeransky and R.~A. Sen'kov.
\newblock Black ring with two angular momenta, hep-th/0612005.

\bibitem{IguchiMishima07}
H. Iguchi and T. Mishima.
\newblock Black diring and infinite nonuniqueness.
\newblock {\em Phys. Rev. {\rm D}}, 75:064018, 2007.

\bibitem{KleKunRad07}
B.~Kleihaus, J.~Kunz, and E.~Radu.
\newblock Rotating nonuniform black string solutions,
\newblock hep-th/0702053.

\bibitem{EmpRea06}
R. Emparan and H.~S. Reall.
\newblock Black rings.
\newblock {\em Class. Quantum Grav.}, 23:R169--R197, 2006.

\bibitem{PodVes98czjp}
J.~Podolsk\'y and K.~Vesel\'y.
\newblock New examples of sandwich gravitational waves and their impulsive
  limit.
\newblock {\em Czech. J. Phys.}, 48:871--878, 1998.

\bibitem{Coleyetal03}
A.~Coley, R.~Milson, N.~Pelavas, V.~Pravda, A.~Pravdov\'a, and R.~Zalaletdinov.
\newblock Generalizations of \pp-wave spacetimes in higher dimensions.
\newblock {\em Phys. Rev. {\rm D}}, 67:104020, 2003.

\bibitem{LewPaw05}
J.~Lewandowski and T.~Pawlowski.
\newblock Quasi-local rotating black holes in higher dimension: geometry.
\newblock {\em Class. Quantum Grav.}, 22:1573--1598, 2005.

\bibitem{Hollandsetal06}
S.~Hollands, A.~Ishibashi, and R.~M. Wald.
\newblock A higher dimensional stationary rotating black hole must be
  axisymmetric.
\newblock {\em Commun. Math. Phys.}, 271:699--722, 2007.

\bibitem{Ficken39}
F.~A. Ficken.
\newblock The {R}iemannian and affine differential geometry of product-spaces.
\newblock {\em Ann. Math.}, 40:892--913, 1939.

\bibitem{FreRub80}
P.~G.~O. Freund and M.~A. Rubin.
\newblock Dynamics of dimensional reduction.
\newblock {\em Phys. Lett. {\rm B}}, 97:233--235, 1980.

\bibitem{CarDiaLem04}
V.~Cardoso, \'{O}. J.~C. Dias, and J.~P.~S. Lemos.
\newblock Nariai, {B}ertotti-{R}obinson and anti-{N}ariai solutions in higher
  dimensions.
\newblock {\em Phys. Rev. {\rm D}}, 70:024002, 2004.

\bibitem{RamosVazJMP03}
M.~P.~M. Ramos and E.~G. L.~R. Vaz.
\newblock Double warped space–times.
\newblock {\em J. Math. Phys.}, 44:4839--4865, 2003.

\bibitem{CardaC93}
J.~Carot and J.~da~Costa.
\newblock On the geometry of warped spacetimes.
\newblock {\em Class. Quantum Grav.}, 10:461--482, 1993.

\bibitem{PleSta68}
J.~Pleba\'nski and J.~Stachel.
\newblock Einstein tensor and spherical symmetry.
\newblock {\em J. Math. Phys.}, 9:269--283, 1979.

\bibitem{HorRos97}
G.~T. Horowitz and S.~F. Ross.
\newblock Properties of naked black holes.
\newblock {\em Phys. Rev. {\rm D}}, 57:1098--1107, 1998.

\bibitem{PodOrt06}
J.~Podolsk\'y and M.~Ortaggio.
\newblock {R}obinson-{T}rautman spacetimes in higher dimensions.
\newblock {\em Class. Quantum Grav.}, 23:5785--5797, 2006.

\bibitem{Ortaggio07}
M.~Ortaggio.
\newblock Higher dimensional spacetimes with a geodesic, shearfree, twistfree
  and expanding null congruence, gr-qc/0701036.

\end{thebibliography}
\end{document}